\documentclass[12pt]{iopart}

\expandafter\let\csname equation*\endcsname\relax
\expandafter\let\csname endequation*\endcsname\relax

\pdfoutput=1

\usepackage[T1]{fontenc}
\usepackage{graphicx} 
\usepackage{subcaption}
\usepackage{multirow}
\usepackage{color}
\usepackage[backref=none]{hyperref}
\hypersetup{
  colorlinks,
  citecolor=red,
  filecolor=black,
  linkcolor=blue,
  urlcolor=magenta
}
\usepackage{amssymb,amsfonts,amsmath}
\usepackage[mathlines]{lineno}
\newcommand*\patchAmsMathEnvironmentForLineno[1]{%
  \expandafter\let\csname old#1\expandafter\endcsname\csname #1\endcsname
  \expandafter\let\csname oldend#1\expandafter\endcsname\csname end#1\endcsname
  \renewenvironment{#1}%
     {\linenomath\csname old#1\endcsname}%
     {\csname oldend#1\endcsname\endlinenomath}}%
\newcommand*\patchBothAmsMathEnvironmentsForLineno[1]{%
  \patchAmsMathEnvironmentForLineno{#1}%
  \patchAmsMathEnvironmentForLineno{#1*}}%
\AtBeginDocument{%
\patchBothAmsMathEnvironmentsForLineno{equation}%
\patchBothAmsMathEnvironmentsForLineno{align}%
\patchBothAmsMathEnvironmentsForLineno{flalign}%
\patchBothAmsMathEnvironmentsForLineno{alignat}%
\patchBothAmsMathEnvironmentsForLineno{gather}%
\patchBothAmsMathEnvironmentsForLineno{multline}%
}

\usepackage{cite}

\newcommand{\kappas}{\kappa_{\textsc{s}}}
\newcommand{\kappad}{\kappa_{\textsc{d}}}
\newcommand{\cL}[1][{}]{\mathcal{L}_{#1}}
\newcommand{\tc}{t}

\newcommand{\avgeq}[1]{\left \langle #1 \right \rangle_{\textsc{eq}}}
\newcommand{\avgneq}[1]{\left \langle #1 \right
  \rangle_{\textsc{neq}}}
\newcommand{\braketH}[2]{\langle #1\vert#2 \rangle_\mathcal{H}}
\newcommand{\ketH}[1]{\vert #1 \rangle_\mathcal{H}}
\newcommand{\Peq}{P_\textsc{eq}}
\newcommand{\deltad}{\delta_\mathrm{D}}
\newcommand{\hL}{\widehat{L}}
\newcommand{\htau}{\widehat{\tau}}
\newcommand{\Psist}{\Psi_\textsc{{st}}}
\newcommand{\La}[1]{L_{#1}^{({1}/{2})}}
\newcommand{\gammars}[1]{\gamma^{(#1)}}
\newcommand{\Jh}{J_{\textsc{h}}}
\newcommand{\dddots}{...}

\numberwithin{equation}{section} 
\numberwithin{figure}{section} 

\allowdisplaybreaks[3]

\begin{document}

\title[Dynamical contribution to the heat conductivity \dots]
{Dynamical contribution to the heat conductivity in
  stochastic energy exchanges of locally confined gases} 

\author{Pierre Gaspard and Thomas Gilbert}
\address{
  Center for Nonlinear Phenomena and Complex Systems,
  Universit\'e Libre  de Bruxelles, C.~P.~231, Campus Plaine, B-1050
  Brussels, Belgium
}

\date{Version of \today}

\begin{abstract}
  We present a systematic computation of the heat conductivity of
  the Markov jump process modeling the energy exchanges in an array
  of locally confined hard spheres at the conduction threshold.  Based on
  a variational formula [Sasada M 2016, {\it Thermal conductivity for
    stochastic energy exchange models}, arXiv:1611.08866], explicit upper
  bounds on the conductivity are derived, which exhibit a rapid
  power-law convergence towards an asymptotic value. We thereby
  conclude that the ratio of the heat conductivity to the energy
  exchange frequency deviates from its static contribution by a small
  negative correction, its dynamic contribution, evaluated to be
  $-0.000\,373$ in dimensionless units. This prediction is
  corroborated by kinetic Monte Carlo simulations which were
  substantially improved compared to earlier results.
\end{abstract}

\submitto{J. Stat. Mech. Theor. Exp.}

\ead{
  \mailto{pierre.gaspard@ulb.ac.be},
  \mailto{thomas.gilbert@ulb.ac.be}
}


\section{Introduction}

Understanding the transport properties of many-body dynamical systems
remains one among statistical physics' greatest challenges.  Much
effort has thus been devoted to deriving Fourier's law of heat
conduction starting from a microscopic setup.  Drawing upon the analogy
with the problem of diffusion in periodic billiard tables
\cite{Bunimovich:1981p479, Bunimovich:1991p47, Chernov:1994p7678}, 
high-dimensional billiard systems were proposed to investigate heat
transport \cite{Bunimovich:1992p621}. Such billiards, which can
be considered intermediate between the gas of hard balls and the 
periodic Lorentz gas, are designed so as to prevent hard balls from
changing positions in a periodic array of confining cells while letting them
interact pairwise. In a regime of rare interactions, i.e.~the limit
such that the binary collisions transporting energy are much less frequent
than energy-conserving wall collision events, it was argued that
the dynamics of energy exchanges between neighbouring hard balls can
be mapped onto a stochastic model \cite{Gaspard:2008PRL101,
  Gaspard:2008NJP3004, Gaspard:200811P021, Gaspard:2009P08020,
  Gaspard:2012p26117, Balint:2016Ballpiston}. This limit is reached
for a critical geometry of the billiard whereby the system undergoes a
transition from thermal conductor to insulator. The stochastic model
stemming from the separation of the two timescales near the critical
geometry is a Markov jump process for the local energy variables,
which lends itself to a systematic derivation of Fourier's law. Thus
the necessary spectral gap was obtained in reference
\cite{Sasada:2015Gap}; see also reference
\cite{Grigo:2012MixingRate}. These results ultimately make possible  
the actual determination of the coefficient of heat conductivity of
the billiard model at the conduction threshold.  

In our previous works \cite{Gaspard:2008PRL101,  Gaspard:2008NJP3004, 
  Gaspard:200811P021, Gaspard:2009P08020, Gaspard:2012p26117},
as well as in reference \cite{Gilbert:2008PRL101200601}, it was shown
that heat conductivity can be expressed in terms of the frequency of
binary collisions responsible for energy exchanges.  Moreover,
numerical results and theoretical considerations led us to conjecture
that the dimensionless ratio between the heat conductivity and the
frequency of binary collisions should be equal to  unity at the
conductor to insulator threshold.  However, recent  work by Sasada
\cite{Sasada:2016Thermal}, relying upon a transposition of Spohn's
variational formula for diffusion coefficients of stochastic lattice
gases \cite{Spohn1990:1227, Spohn:1991book} to the heat conductivity
of stochastic energy exchange models, proved that this conjecture
cannot hold; the heat conductivity is in fact the sum of two parts:
the static contribution, which is identical, up to a unit length
squared, to the energy exchange frequency, and the dynamic
contribution, which, however small it turns out to be, is strictly
negative. The variational formula, which is the main focus of this
work, was initially derived by Varadhan
\cite{varadhan:1994NonlinearII} in the context of a non-gradient
Ginzburg-Landau model at infinite temperature and has since developed
into a cornerstone of the framework for analyzing non-gradient models; 
see e.g.~reference~\cite{Varadhan:1997p282}.    

In the present paper, we set out to demonstrate that an explicit
calculation of the dimensionless ratio between the heat conductivity
and the energy exchange frequency can be achieved 
using the variational formula. Our method relies on the
use of multivariate polynomial test functions of increasing number of
variables and degrees to compute explicit upper bounds on the heat
conductivity. The rapid power-law convergence towards an asymptotic
value yields a narrow confidence interval for this quantity. 

The predicted value thus obtained is in fact within the error bounds
of the (inconclusive) numerical findings reported in reference
\cite{Gaspard:2009P08020} and therefore calls for a careful revision
of our kinetic Monte Carlo simulations and the procedure by which the
measurement of the heat conductivity is performed so as to improve its
precision and confirm our new findings. We thereby show that the negative
dynamic contribution to the heat conductivity inferred from the
variational formula is in agreement with the value obtained from
kinetic Monte Carlo simulations. This result confirms Sasada's
conclusion \cite{Sasada:2016Thermal} and provides a prediction of the heat
conductivity accurate to within at least six significant digits. 

The paper is organized as follows. In \sref{sec:Overview}, the
characterization of heat transport in many-body billiard systems with
caged hard balls is reviewed from both microscopic and macroscopic
considerations in subsection \ref{sec:Micro}. In
subsection \ref{sec:Meso}, its transposition to stochastic energy 
exchange systems at the mesoscopic level is given. Within this
framework, the variational formula is established in
\sref{sec:varformula}.  In particular, the application of the method
to the stochastic energy exchanges induced by locally confined hard
spheres is discussed in subsection \ref{sec:Application}.  In
subsection \ref{sec:Calculation-(r=2)}, the implementation of the
variational formula is described for bivariate trial functions. Its
extension to multivariate trial functions and its extrapolation to
infinite-dimensional functions are presented in subsection
\ref{sec:Calculation-(r>2)}.  A comparison of the results thus 
obtained with kinetic Monte Carlo simulations is established in
\sref{sec:Simulations}.  Section~\ref{sec:Conclusion} concludes the 
paper.

\section{Energy transport in locally confined gases}
\label{sec:Overview}

\subsection{\label{sec:Micro} From microscopic dynamics to macroscopic
  fields} 

In the models described in references~\cite{Gaspard:2008PRL101,
  Gaspard:2008NJP3004, Gaspard:200811P021, Gaspard:2009P08020,
  Gaspard:2012p26117}, we considered the motion of hard
$D$-dimensional balls, $D\geq 2$, on a periodic $d$-dimensional array
of cells, $1 \leq d \leq D$, undergoing elastic collisions with
locally confining hard walls as well as among neighbours. While the
choice of the values of these two dimensions is mostly a matter of
convenience, the distinction between the two is important. 
Unless otherwise stated, we will assume $D=3$ and $d=1$.

\begin{figure}[htb]
  \centering{
    \includegraphics[width=0.9\textwidth]{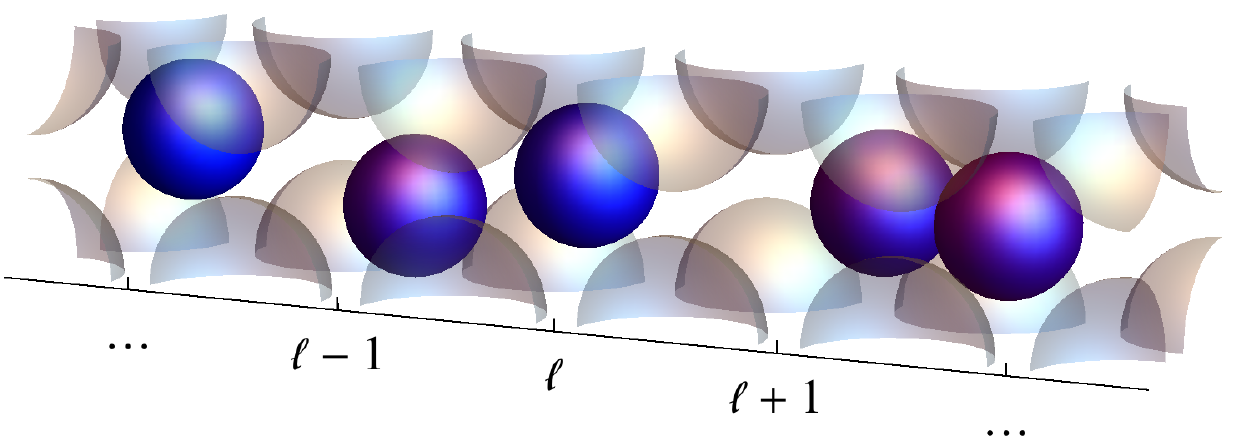}
  }
  \caption{Schematic view of a billiard table composed of locally
    confined hard spheres. Mass transport is blocked by the geometry,
    but energy is transported through binary collisions among
    neighbouring hard balls, such as with the right-most pair.}
  \label{fig:billiard}
\end{figure}

Such a three-dimensional model on a one-dimensional lattice is
schematically depicted in \fref{fig:billiard}. In this example,
identical hard spheres are randomly placed along a 
one-dimensional periodic array $\cL[N] = \{-(N-1)/2, \dots, (N-1)/2\}$
of $N$ small cavities composed of fixed spherical obstacles located at
the corners of a cube of side $\ell$. To fix notations, we let the
center of each cavity be located at $x_{l}=l \ell$. Each such cavity
contains a single hard sphere which rattles around in it, unable to
escape. The geometry is however chosen so as to allow neighbouring
hard spheres to collide with each other. 

In every cell, the motion of hard spheres is a succession of free
flights interrupted by specular reflections which consist of
two types of events: (i) wall collision events, through which a single
ball reflects elastically off the spherical boundary of the cavity
and does not change its energy, or (ii) binary collision
events, which occur when two neighbouring balls collide with each
other, thereby exchanging energy. The billiard geometry must verify
specific conditions for such a regime to take place; see 
references~\cite{Gaspard:2008NJP3004, Gaspard:2012p26117}.

The system therefore consists in a gas of $N$ hard spheres whose
spatial ordering is preserved by the dynamics. A current of energy may
however be induced by the binary collision events, bringing about
conduction of heat across the system.

For such a deterministic system, the local conservation of energy
$\epsilon_{l}$ in cell $l = (l_1,\dots,l_{d})$ can
be expressed as 
\begin{equation} 
  \dot \epsilon_{l}(t) = \sum_{i = 1}^{d}
  \left[
    \mathcal{J}_{l-1_{i},l} (t) - \mathcal{J}_{l,l+1_{i}} (t)
    \right]\,,
  \label{eq:det-energy-cons} 
\end{equation} 
in terms of the deterministic instantaneous currents
$\mathcal{J}_{l,l+1_{i}}(t)$, where $1_{i}$ is the vector whose $d$
components are all $0$ but for the $i$\textsuperscript{th} which is
$1$. In the case of billiards, these currents have the form 
\begin{equation} 
  \label{eq:detcurrents}
  \mathcal{J}_{l,l+1_{i}}(t) = 
  \sum_{n} \eta_{l,l+1_{i}}(\tc) \, \deltad(t-\tc^{(l,l+1_{i})}_{n}) \, , 
\end{equation} 
where $\eta_{l,l+1_{i}}(\tc)$ denotes the energy exchanged
through the collision between balls in cells $l$ and $l+1_{i}$ at
time $t$, $\{\tc^{(l,l+1_{i})}_{n}\}_{n\in\mathbb{N}}$ is the sequence of 
all successive collision times corresponding to binary collisions
between balls $l$ and $l+1_{i}$ and the symbol  $\deltad$ stands for
the Dirac delta distribution. Integrating this quantity over time and
summing over the lattice dimensions yields the kinetic energy of the
$l^{\mathrm{th}}$ ball,  
\begin{equation}
  \epsilon_{l}(t)-\epsilon_{l}(0) = 
  \sum_{i=1}^{d} 
  \int_0^t \rmd t' 
  \left[
    \mathcal{J}_{l-1_{i},l}(t') -
    \mathcal{J}_{l,l+1_{i}} (t')
  \right]
    \, .
  \label{eq:ej} 
\end{equation}

The connection with the macroscopic description is established by
introducing the local temperature in terms of the kinetic energy
averaged over some nonequilibrium statistical ensemble, 
\begin{equation}
  \label{eq:localTemp}
  T(x_{l},t) = \frac{2}{D} \avgneq{\epsilon_{l}(t)}
  \,,
\end{equation} 
where $\avgneq{\cdot}$ denotes the time-dependent average with respect
to the nonequilibrium statistical ensemble\footnote{Throughout the 
  paper, Boltzmann's constant is taken to be unity, $k_{\mathrm{B}}
  \equiv 1$, so that temperatures and energies are expressed in the
  same units.}. Similarly, the local energy and energy current
densities, with respect to cells of (hyper)cubic volume $\ell^{d}$ in a
$d$-dimensional lattice, can be defined as   
\begin{equation}
  \label{eq:jemacro}
  \begin{split}
    e(x_{l}, t) &= \ell^{-d} \, \avgneq{\epsilon_{l}(t)} \, , \\
    j_{e,i}(x_{l}, t) &= \ell^{-d+1} \, \tfrac{1}{2}
    \avgneq{ 
        \mathcal{J}_{l-1_{i},l}(t) + \mathcal{J}_{l,l+1_{i}}(t)
    } \, ,
  \end{split}
\end{equation}
in terms of the deterministic currents \eqref{eq:detcurrents}. The
heat capacity is thus given by    
\begin{equation}
c = \frac{\partial e}{\partial T} =  \frac{D}{2} \ell^{-d}\, ,
\end{equation}
under the assumption of local equilibrium. 

If it exists, the heat conductivity enters at the macroscopic level,
the expression of Fourier's law, obtained in the hydrodynamic scaling
limit where $\ell\to0$, 
\begin{equation}
  \label{eq:Fourier}
  j_{e}(x,t) = -\kappa(T(x,t)) \, \mathrm{grad}\,T(x,t)\,,
\end{equation}
where $j_{e}$ is the vector of components $j_{e,i}$, $i=1,\dots,d$,
defined in equation \eqref{eq:jemacro}.
Moreover, since the energy current density $j_{e}$ obeys the
local conservation law, 
\begin{equation}
  \label{eq:energyflux}
  \partial_t \, e(x,t) + \mathrm{div}\, j_{e}(x,t) = 0 \, ,
\end{equation}
the heat equation can be written as
\begin{equation}
  \label{eq:FourierTemp}
  \partial_t T(x,t) = \mathrm{div}  
  \left[\chi(T(x,t)) \, \mathrm{grad} \, T(x,t)\right]
  \,,
\end{equation}
in terms of the thermal diffusivity
\begin{equation}
  \label{eq:diffusivity}
  \chi(T) = c^{-1}\kappa(T) = \frac{2\, \ell^d}{D} \, \kappa(T) \, .
\end{equation}
The diffusivity and conductivity therefore differ in their
units. Only the latter quantity will be referred to below.

In deterministic systems of $N$ balls with position coordinates
$q_{l}$ and energies $\epsilon_{l}$ in a volume $V_{N}$, the heat
conductivity is in general given by  
Helfand's formula \cite{Helfand:1960p18123},
\begin{equation} 
  \kappa(T) = \lim_{t\to \infty}\lim_{N\to\infty}
  \frac{1}{2t V_{N} T^2} \avgeq{\left[G(t)-G(0)\right]^2}\,,
  \label{eq:Helfand} 
\end{equation} 
where $G(t) = \sum_{l = 1}^{N} q_{l} \, \epsilon_{l}$ is
Helfand's moment associated with energy, $\avgeq{\cdot}$ denotes
the average over the equilibrium statistical ensemble at the
temperature $T$ and we assume the ratio $V_{N}/N$ is fixed as $N$
increases. For periodic arrays of cells of volume $\ell^{d}$, each
containing a single ball, we have $V_{N} = N \ell^{d}$. 
Although the Green-Kubo formulation of the transport
coefficients is often favoured over Helfand's, the two are in fact
equivalent; see reference~\cite{Helfand:1960p18123}.

In $(d=1)$-dimensional chains, such as depicted in
\fref{fig:billiard}, the confining geometry allows to simplify
Helfand's moment to  
\begin{equation} 
  G(t) = \sum_{l\in\cL[N]} l \ell \,
  \epsilon_{l}(t)\,,
  \label{eq:G} 
\end{equation} 
which holds irrespective of the spatial dimension $D$ of the
underlying dynamics.  For more general geometries, $\epsilon_{l}$ is
to be interpreted as the energy associated with the gas (of one or
more particles) trapped in cell $l$. 

In the large system-size limit, using translation invariance of energy
correlations and the local conservation of
energy~\eqref{eq:det-energy-cons}, it is possible to transform
\eqref{eq:Helfand} with Helfand's moment \eqref{eq:G}
to
\begin{equation} 
  \kappa(T) =
  -\frac{\ell}{4T^2} \lim_{t\to\infty} \frac{1}{t} \sum_{l\in\mathbb{Z}} l^2 
  \avgeq{
    \left[ \epsilon_{l}(t) - \epsilon_{l}(0) \right] 
    \left[ \epsilon_{0}(t) - \epsilon_{0}(0) \right]}\,,
  \label{eq:kappa} 
\end{equation} 
which applies to an infinite system; see also
\cite[equation~(3.1)]{Sasada:2016Thermal}. As explained in \sref{sec:Meso},
the contributions to this equation  can be conveniently 
separated into static and dynamic correlations, so one can write
\begin{equation}
  \label{eq:kappasd}
  \kappa(T) \equiv \kappas(T) + \kappad(T)\,.
\end{equation}
Whereas the static contribution, $\kappas(T)$, is easily
computed, the dynamic contribution, $\kappad(T)$, is more elusive. 
Its computation will be the focus of section \ref{sec:varformula}.

\subsection{\label{sec:Meso}Mesoscopic description}

As argued previously~\cite{Gaspard:2008PRL101, Gaspard:2008NJP3004,
  Gaspard:200811P021, Gaspard:2009P08020, Gaspard:2012p26117}, the
process of heat transport in the billiards described above reduces to
a stochastic energy-exchange process in a regime of 
rare interactions, which occurs when the frequency of binary collisions
is much smaller than the frequency of wall collisions. Whereas the
former governs the timescale of energy exchanges, the latter
randomizes the degrees of freedom not relevant to energy transport,
i.e.~ the positions and velocity directions. In other words, the
separation between the two timescales induces averaging of these
degrees of freedom.

Setting $d=1$, an energy configuration $\{\epsilon_{l}\}_{l\in\cL[N]}$
thus evolves according to a Markov jump process. On the one hand, the
time evolution of the probability density $P \equiv
P(\{\epsilon_{l}\}_{l\in\cL[N]}, t)$ is specified by the master  equation, 
\begin{equation}
  \partial_t P = \hL^\dagger \, P = \sum_{l\in\cL[N]} \hL_l^\dagger P \, ,
  \label{eq:master} 
\end{equation} 
with the local energy-exchange operator, 
\begin{multline} 
  \hL_l^\dagger
  P(\dots,\epsilon_{l},\epsilon_{l+1},\dots) = 
  \int_{-\epsilon_{l+1}}^{\epsilon_{l}} \rmd\eta 
  \Big[
  W(\epsilon_{l} - \eta, \epsilon_{l+1} + \eta \,\vert\,
  \epsilon_{l}, \epsilon_{l+1})\,
  P(\dots, \epsilon_{l} - \eta, \epsilon_{l+1} +\eta, \dots) 
  \cr 
  -
  W(\epsilon_{l},\epsilon_{l+1}\,\vert\,
  \epsilon_{l} - \eta, \epsilon_{l+1} + \eta)
  \,
  P(\dots,\epsilon_{l},\epsilon_{l+1},\dots)\Big] ,
\end{multline} 
whose structure is the usual difference between gain and
loss terms, defined in terms of a stochastic kernel $W$ to be
specified below. On the other hand, observables $A \equiv
A(\{\epsilon_{l}\}_{l\in\cL[N]}, t )$ evolve in time under the action
of the adjoint operator, $\hL$, which takes the somewhat simpler form: 
\begin{equation}
  \partial_t A = \hL \, A = \sum_{l\in\cL[N]} \hL_l A \, ,
  \label{eq:L} 
\end{equation} 
with 
\begin{multline} 
  \hL_l A(\dots,\epsilon_{l},\epsilon_{l+1}, \dots)
  =
    \int_{-\epsilon_{l+1}}^{\epsilon_{l}} \rmd\eta 
    \, W(\epsilon_{l},\epsilon_{l+1}\vert
    \epsilon_{l}-\eta,\epsilon_{l+1}+\eta) 
    \cr
    \times
    \Big[
    A(\dots,\epsilon_{l}-\eta,\epsilon_{l+1}+\eta,\dots)
    - A(\dots,\epsilon_{l},\epsilon_{l+1},\dots)\Big] 
    \,.
    \label{eq:L_l} 
\end{multline} 
Assuming periodic boundary conditions, the terms $l = (N-1)/2$
in \eqref{eq:master} and \eqref{eq:L} couple the right-most cell
$(N-1)/2$ with the left-most one $-(N-1)/2$.

For future reference, note that, in infinite-size systems, the
natural equilibrium distribution for $D=3$ is the canonical
probability distribution, 
\begin{equation} 
  \Peq(\dots,\epsilon_{l},\epsilon_{l+1},\dots) =
  \prod_{l\in\cL[N]} 
  \left[\frac{2\beta}{\sqrt{\pi}} \sqrt{\beta\epsilon_{l}} \,
    \rme^{-\beta\epsilon_{l}} \right] .
  \label{eq:P_eq} 
\end{equation} 
For general $D$, it is given by the product of Gamma distributions of
shape parameter $\tfrac{D}{2}$ and scale parameter specified by the
temperature $T = \beta^{-1}$, such that $\lim_{N\to\infty} N^{-1} 
\sum_{l\in\cL[N]} \epsilon_{l} = \tfrac{D}{2}T$. This is a stationary
solution of the master equation~\eqref{eq:master}. 

In this stochastic description, the local conservation of energy
becomes 
\begin{equation}
  \partial_t \, \avgneq{\epsilon_{l}} = 
  \avgneq{ j(\epsilon_{l-1},\epsilon_{l})} - 
  \avgneq{ j(\epsilon_{l},\epsilon_{l+1})} 
\end{equation} 
where $\avgneq{\cdot}$ now denotes the statistical average
with respect to the time-dependent nonequilibrium probability
distribution~$P$ and the local average current is defined to be: 
\begin{equation} 
  j(\epsilon_{l},\epsilon_{l+1}) = 
  \int \rmd\eta \, \eta \, W(\epsilon_{l},\epsilon_{l+1}\vert
  \epsilon_{l}-\eta,\epsilon_{l+1}+\eta)\,.
  \label{eq:loc_av_curr} 
\end{equation} 
This observable is antisymmetric under the permutation of
its arguments,
\begin{equation}
  j(\epsilon_{l},\epsilon_{l+1}) =-j(\epsilon_{l+1},\epsilon_{l}) \, .
  \label{eq:j-anti} 
\end{equation} 

In the stochastic description, contrary to what
equation~\eqref{eq:ej} of the deterministic description would suggest,
the change of energy at site $l$ in time $t$ is not simply given by
the time integration of the local average
currents~\eqref{eq:loc_av_curr}. Rather, an additional contribution
must be taken into consideration, given by a 
martingale\footnote{A martingale is defined by the property that its
  expectation value conditioned on the knowledge of the whole process 
until the time $t'<t$ is equal to the value of the martingale at time
$t'$ \cite{Pinsky:2011}: 
\[
  \mathbb{E}\left[ M_{l}(t) \vert \{\epsilon_k(t'')\}_{k\in\cL[N]}, t''<t'
  \right] = M_{l}(t')\,, \qquad \forall t' <t \, .  
\]
This property has been established for stochastic lattice gases
\cite{Spohn1990:1227, Spohn:1991book} as well as for stochastic energy
exchange models \cite{Basile:2009p287}.  
} 
$M_{l}(t)$, 
\begin{equation}
  \epsilon_{l}(t)-\epsilon_{l}(0) = M_{l}(t) + \int_0^t \rmd t'
  j\left[\epsilon_{l-1}(t'),\epsilon_{l}(t')\right] - \int_0^t \rmd t'
  j\left[\epsilon_{l}(t'),\epsilon_{l+1}(t')\right] \, .
  \label{eq:eMj} 
\end{equation} 
In this expression the left-hand side is the exact change of
energy at site $l$ in time $t$. The integrals on the right-hand side are, 
however, carried out over the currents~\eqref{eq:loc_av_curr}, which
correspond to averaged quantities. By contrast, the actual succession
of random energy jumps in time $t$ displays fluctuations about these
average currents. The martingale $M_{l}(t)$ thus represents the difference
between the actual change of energy in time $t$ and the corresponding
time integrals of the local average currents. Considering the
differential of this expression, $\rmd \epsilon_{l}(t) =
j\left[\epsilon_{l-1}(t),\epsilon_{l}(t)\right] \, \rmd t 
- j\left[\epsilon_{l}(t),\epsilon_{l+1}(t)\right] \, \rmd t + \rmd
M_{l}(t)$, we see that $\rmd M_{l}(t)$ is akin to a Langevin noise term.

Since martingales have independent increments, $M_{l}$ satisfies the
following property 
\begin{equation} 
  \label{eq:avgMM}
  \avgeq{M_{l}(t) M_0(t)} =
  -2 t \avgeq{ \epsilon_{l}
    \left[ j(\epsilon_{-1},\epsilon_{0}) - 
      j(\epsilon_{0},\epsilon_{1}) \right] } ;
\end{equation} 
see \cite{Sasada:2016Thermal}. Solving equation \eqref{eq:eMj} for $M_{l}(t)$,
the left-hand side of equation \eqref{eq:avgMM} can be expressed in
terms of the sum of three terms, one involving correlations between
energy changes, as they appear on the right-hand side  of equation
\eqref{eq:kappa}, a second term involving correlations between energy
changes and time-integrals of the local average currents, and a third one
involving self-correlations of time-integrals of the local average
currents. As pointed out by Spohn \cite{Spohn1990:1227,
  Spohn:1991book}, the equilibrium averages of the cross-terms between 
the change of the  conserved quantity, given by the left-hand side of
equation \eqref{eq:eMj}, and the time integral of the local 
average currents, as on the right-hand side of the same equation,
must vanish because the former is odd under time 
reversal while the latter is even. Therefore, after summing
\eqref{eq:avgMM} over $l\in\mathbb{Z}$ and  transforming the
correlation functions using translation invariance, we can multiply
the resulting expression by $-\ell/(4 T^{2} \, t)$ and take the limit
$t\to\infty$, as in equation \eqref{eq:kappa}, to obtain the
expression of the heat conductivity\footnote{Henceforth, we assume the 
  array $\cL$ to have infinite extension on both sides so that $\cL
  \equiv \cL[\infty] \sim \mathbb{Z}$.} 
\begin{equation} 
  \kappa(T) = \tfrac{1}{2}\ell\,T^{-2} 
  \avgeq{ (\epsilon_{0}-\epsilon_{1}) j(\epsilon_{0},\epsilon_{1})} 
  - \ell\, T^{-2} \sum_{l\in\mathbb{Z}} \int_0^\infty \rmd t 
  \avgeq{
    j(\epsilon_{0},\epsilon_{1})\, \rme^{\hL t}
    j(\epsilon_{l},\epsilon_{l+1})},
  \label{eq:kappa2} 
\end{equation} 
where we recall $\hL=\sum_{l\in\mathbb{Z}} \hL_l$ is the
time-evolution generator \eqref{eq:L}-\eqref{eq:L_l} for the
observables. Remark here that a necessary condition for the time
integral to converge is that $\hL$ be non-positive definite. 

The first term on the right-hand side of
equation~\eqref{eq:kappa2} is identified as the static contribution to
the heat conductivity in equation~\eqref{eq:kappasd},  
\begin{equation} 
  \kappas(T) = \tfrac{1}{2}\ell\, T^{-2}
  \avgeq{ (\epsilon_{0}-\epsilon_{1}) j(\epsilon_{0},\epsilon_{1})} 
  = \tfrac{1}{2}\ell\, T^{-2} \avgeq{ h(\epsilon_{0},\epsilon_{1}) }
  = \ell\,  \avgeq{ \nu(\epsilon_{0},\epsilon_{1}) } ,
  \label{eq:kappa_s} 
\end{equation} 
where
\begin{align}
  \nu(\epsilon_{0},\epsilon_{1}) 
  &
    = \int \rmd\eta \,
    W(\epsilon_{0},\epsilon_{1}\vert \epsilon_{0}-\eta,\epsilon_{1}+\eta)\,
    , 
    \label{eq:nu}
  \\ 
  h(\epsilon_{0},\epsilon_{1}) 
  &
    = \int \rmd\eta \, \eta^2
    \, W(\epsilon_{0},\epsilon_{1}\vert
    \epsilon_{0}-\eta,\epsilon_{1}+\eta)\,,     
    \label{eq:h}
\end{align} 
are the zeroth and second moments of the stochastic kernel. 
The former is nothing but the mean frequency of binary
collisions for the corresponding energy pair.  The second term on the
right-hand side of equation~\eqref{eq:kappa2} is therefore identified
as the dynamic contribution, 
\begin{equation}
  \label{eq:kappad}
  \kappad(T) 
  = 
  -\ell\,   T^{-2} \sum_{l\in\mathbb{Z}} \int_0^\infty \rmd t 
  \avgeq{
    j(\epsilon_{0},\epsilon_{1})\, \rme^{\hL t}
    j(\epsilon_{l},\epsilon_{l+1})}\,.
\end{equation}

Before we turn to the characterization of this second term in
\sref{sec:varformula}, we note that one can formally write: 
\begin{equation} 
  \label{eq:intL}
  \int_0^{\infty} \rmd t\, \rme^{\hL t} = - \hL^{-1} \, ,
\end{equation} 
which assumes the operator $\hL$ causes relaxation after
long enough times.  Accordingly, equation~\eqref{eq:kappa2} can be
rewritten as the statistical average of the heat current
$j(\epsilon_{0},\epsilon_{1})$, 
\begin{equation} 
  \kappa(T) = -\avgeq{ j(\epsilon_{0},\epsilon_{1}) \, \Psist},
  \label{eq:kappa-stst} 
\end{equation} 
where
\begin{equation} 
  \Psist \equiv \Psist(\{\epsilon_{l}\}_{l\in\mathbb{Z}})
  = \tfrac{1}{2}\ell\, T^{-2} (\epsilon_{1}-\epsilon_{0}) -
  \ell\, T^{-2} \sum_{l\in\mathbb{Z}} \hL^{-1} j(\epsilon_{l},\epsilon_{l+1}) \, ,
  \label{eq:stst} 
\end{equation}
is an infinite-dimensional function to be interpreted in
terms of the first-order expansion of a nonequilibrium steady state in
powers of its local temperature gradient.

Indeed, consider the transposition of equation \eqref{eq:Fourier} to a
nonequilibrium steady state with temperature profile $T(x_{l})$,
$x_{l} = l\,\ell$, $l\in\cL[N]$, resulting from the presence of heat baths
at different temperatures at the system boundaries\footnote{Let $x
  \equiv x_{l}$. Due to
  the square-root dependence of the conductivity on the temperature,
  the temperature profile is given, asymptotically in $N$, by  
  \[
    T(x) = \left[ \frac{T_{+}^{3/2} + T_{-}^{3/2}}{2} + 
      \left( T_{+}^{3/2} - T_{-}^{3/2} \right) \frac{x}{(N+1)\ell}\right]^{2/3}\,,
  \]
  where $T_{\pm} =T(\pm\tfrac{N+1}{2}\ell)$ are the temperatures of the
  heat baths. Its derivative is therefore proportional to the inverse
  square root of the temperature,
  \[
    \frac{\rmd T}{\rmd x} = \frac{2}{3(N+1)\ell} 
    \frac{T_{+}^{3/2} - T_{-}^{3/2}}{\sqrt{T(x)}}\,.
  \]
  The product of this quantity by $\kappa(T)$ on the right-hand side
  of equation \eqref{eq:Fourier} is a number independent of  $x$,
  which justifies the transposition of equation \eqref{eq:Fourierness}
  in terms of averages of mesoscopic fluctuating quantities analogous
  to equation \eqref{eq:kappa-stst}.
}. 
Assuming $N$ large, we have
\begin{equation}
  \label{eq:Fourierness}
  \kappa(T(x_{l})) = - \left( \frac{\rmd T}{\rmd x_{l}} \right)^{-1}
  j_{e}(x_{l}),
\end{equation}
where the macroscopic current $j_{e}(x_{l})$ should, in analogy with the
second of equations \eqref{eq:jemacro}, be expressed in terms of the
symmetrized sum $\tfrac{1}{2}\avgneq{j(\epsilon_{l-1}, \epsilon_{l}) +
j(\epsilon_{l}, \epsilon_{l+1})}$.
Comparing with equation \eqref{eq:kappa-stst}, we see that the
two-cell marginal of $\Psist$,
\begin{equation}
  \label{eq:Psistmarginal}
  \int \prod_{l\neq 0,\, 1} \rmd \epsilon_{l} \Psist( \dots,
  \epsilon_{0}, \epsilon_{1}, \dots)\,,
\end{equation}
can be identified as the coefficient of the first-order contribution
in the local temperature gradient to the probability density of
the nonequilibrium steady state with respect to the 
equilibrium state \eqref{eq:P_eq}. To be more precise, it is the part
of the nonequilibrium steady state which contributes to the
current. This is, however, not to say that $\Psist$ fully accounts for
the actual density of the nonequilibrium steady state. 

\section{Variational formula \label{sec:varformula}}

\subsection{General framework \label{sec:genframe}}

As shown by Spohn in reference~\cite{Spohn:1991book}, a Hilbert space
$\cal H$ can be introduced for complex functions $f,g$ of the variables
$\{\epsilon_{l}\}_{l\in\mathbb{Z}}$ with the degenerate scalar
product: 
\begin{equation} 
  \label{eq:scalarproduct}
  \braketH{f}{g} \equiv \sum_{k\in\mathbb{Z}} \Big(
    \avgeq{ f^{\star} \htau^{k} g} - \avgeq{f^{\star}} \avgeq{g}
    \Big) \, , 
\end{equation} 
where ${}^\star$ denotes the complex conjugation and $\htau^{k}$ is
the translation operator by $k$ cells to the left, mapping 
$\{\epsilon_{l}\}_{l}$ onto $\{\epsilon'_{l} = \epsilon_{l+k}\}_{l}$.
This scalar product is well defined, i.e.~the sum over $k$ converges,
because the equilibrium distribution 
has the mixing property under the spatial translations
$\{\htau^{k}\}_{k}$ \cite{Spohn:1991book}.  
Let $\hL$ also denote the extension of the operator
\eqref{eq:L_l} to this Hilbert space.  It is proved in
references~\cite{Spohn1990:1227, Spohn:1991book} that  
\begin{equation} 
  \label{eq:infimum} 
  \mathrm{inf}_f\Big(-\braketH{f}{\hL f} - 2 \, \braketH{ j}{ f } \Big) =
  \braketH{j}{\hL^{-1} j } \, ,
\end{equation} 
where the infimum can be taken over real functions $f$.  This result
is obtained by expanding the vectors and the 
operator in a complete basis of the Hilbert space and by taking the
first variation with respect to the function $f$, which leads to 
\begin{equation}
  \braketH{ \delta f}{ \hL f } + \braketH{ \delta f}{ j } = 0 \, .  
\end{equation} 

Formally, this implies that
$\ketH{f} = - \ketH{\hL^{-1} j}$.
Making this substitution in the left-hand side of
equation~\eqref{eq:infimum}, one obtains the right-hand
side. The infimum is thus explicitly realized for the function  
\begin{equation} 
  f_{\textsc{inf}}=-\sum_{k\in\mathbb{Z}}\hL^{-1}\htau^{k} j\,,
  \label{eq:f_inf} 
\end{equation} 
so that $\Psist$ in \eqref{eq:stst} can be written as 
\begin{equation} 
  \Psist = \tfrac{1}{2}\ell\,  T^{-2} (\epsilon_{1}-\epsilon_{0}) +
  \ell\,  T^{-2} f_{\textsc{inf}} \, .
  \label{eq:stst2} 
\end{equation}

Now, from equation~\eqref{eq:intL}, the dynamic contribution to the
heat conductivity \eqref{eq:kappad} can be expressed, in 
terms of the scalar product \eqref{eq:scalarproduct} and using
$\avgeq{ j } \equiv 0$, as 
\begin{align} 
  \label{eq:kappadinf}
  \kappad(T) 
  & = \ell\,  T^{-2} 
    \sum_{k\in\mathbb{Z}} \avgeq{ j \, \hL^{-1} \htau^{k} \, j}
    \,,\cr
  &
    =\ell\,  T^{-2} \braketH{ j }{ \hL^{-1} \, j }
    \,,\cr
  &
    =\ell\,  T^{-2} \mathrm{inf}_f\left(-
    \braketH{ f } {\hL f} - 2 \, \braketH{ j }{ f} 
  \right) \,,
\end{align} 
where the last line follows from equation~\eqref{eq:infimum}. 

Using equation~\eqref{eq:L_l} and the detailed balance condition, 
\begin{equation}
  \label{eq:detailedbalance}
  \Peq(\dddots, \epsilon_{0}, \epsilon_{1},\dddots)
  W(\epsilon_{0}, \epsilon_{1}\vert 
  \epsilon_{0}-\eta, \epsilon_{1}+\eta)
  = 
  \Peq(\dddots, \epsilon_{0} -\eta, \epsilon_{1} + \eta, \dddots)
  W(\epsilon_{0} - \eta, \epsilon_{1} + \eta \vert 
  \epsilon_{0}, \epsilon_{1})
  \,,
\end{equation}
the two following identities are obtained for the quantities appearing
on the left-hand side of equation \eqref{eq:kappadinf}: 
\begin{align} 
  -\braketH{ f }{ \hL f } 
  &=
    \tfrac{1}{2} \avgeq{\int \rmd\eta \, W(\epsilon_{0},\epsilon_{1}\vert
    \epsilon_{0}-\eta,\epsilon_{1}+\eta) \Big( \sum_{k\in\mathbb{Z}} D_{0,1,\eta}
    \htau^{k} f \Big)^2} ,
    \label{eq:f-Lf} 
  \\
    -2 \braketH{ j }{ f }
  & = -2 \avgeq{ j \sum_{k\in\mathbb{Z}} \htau^{k}   f} 
    \,, \cr
  &
    = \avgeq{ \int \rmd\eta \,
    W(\epsilon_{0},\epsilon_{1}\vert \epsilon_{0}-\eta,\epsilon_{1}+\eta) \,
    \eta \, \sum_{k\in\mathbb{Z}} D_{0,1,\eta} \htau^{k} f } ,
  \label{eq:j-f} 
\end{align} 
where we introduced the exchange operator $D_{l,l+1,\eta}$, 
\cite{Spohn:1991book},
\begin{equation} 
  D_{l,l+1,\eta}f \equiv
  f(\dots, \epsilon_{l}-\eta, \epsilon_{l+1}+\eta, \dots) -
  f(\dots, \epsilon_{l}, \epsilon_{l+1}, \dots)\,,
\end{equation} 
which, by convention, is a function of the pair of indices $l$ and
$l+1$ rather than the positions of the corresponding variables. 
We note that equation \eqref{eq:f-Lf} further relies on the identity
$\avgeq{L_{l} f} = 0$, itself a consequence of the detailed balance.

Combining these two results with equation~\eqref{eq:kappadinf} and the
expression of the static contribution, equation~\eqref{eq:kappa_s},
the heat conductivity \eqref{eq:kappasd} is finally expressed as the
solution of the variational formula, 
\begin{equation}
  \label{eq:formula} 
  \kappa(T) = \frac{\ell}{2T^2} \, \mathrm{inf}_f
  \avgeq{ \int \rmd\eta \, W(\epsilon_{0},\epsilon_{1}\vert
    \epsilon_{0}-\eta,\epsilon_{1}+\eta)\Big(\eta + 
    \sum_{k\in\mathbb{Z}} D_{0,1,\eta} \htau^{k} f \Big)^2 } \,,
\end{equation} 
in agreement with  Sasada \cite[equation (A.1)]{Sasada:2016Thermal}.
Moreover, the probability density distribution of the nonequilibrium
steady state contributing to the current is locally given by
equation~\eqref{eq:stst2}, in terms of the function realizing the
infimum. 

An important property is that the solutions~\eqref{eq:f_inf} of the
variational formula are defined up to functions of a single
variable. Indeed, a function $\Delta f = \sum_l \varphi(\epsilon_{l})$
may always be added to the solution~\eqref{eq:f_inf} without changing
the value of the conductivity~\eqref{eq:kappa-stst}. This is so
because the antisymmetry~\eqref{eq:j-anti} of the local average
current implies  
\begin{align}
  \label{eq:nosinglevarfunction}
  \avgeq{ j(\epsilon_{0},\epsilon_{1}) \sum_l \varphi(\epsilon_{l})} 
  &
    =\sum_{l\leq -1} \underbrace{
    \avgeq{j(\epsilon_{0},\epsilon_{1})}}_{\, = \, 0}
    \avgeq{\varphi(\epsilon_{l})} + 
    \avgeq{ j(\epsilon_{0},\epsilon_{1}) \varphi(\epsilon_{0})}
    \cr
  &\quad
    + \avgeq{ j(\epsilon_{0},\epsilon_{1}) \varphi(\epsilon_{1})} +
    \sum_{2\leq l} 
    \underbrace{ \avgeq{ j(\epsilon_{0},\epsilon_{1})} }_{\, = \, 0}
    \avgeq{ \varphi(\epsilon_{l}) } 
    \,,
    \cr
  &
    = \avgeq{ j(\epsilon_{0},\epsilon_{1}) \varphi(\epsilon_{0}) }
    - \avgeq{ j(\epsilon_{1},\epsilon_{0}) \varphi(\epsilon_{1}) }
    \,,
    \cr
  &
    = \avgeq{ j(\epsilon_{0},\epsilon_{1}) \varphi(\epsilon_{0}) }
    - \avgeq{ j(\epsilon_{0},\epsilon_{1}) \varphi(\epsilon_{0}) }
    \,,
    \cr
  &
    =0 \, .  
\end{align}
Furthermore and by the same token, only antisymmetric functions of
$\epsilon_{0}$ and $\epsilon_{1}$ can produce non-trivial
contributions to the variational formula. Solutions \eqref{eq:f_inf}
are therefore defined up to symmetric functions of these arguments.

\subsection{Application to the stochastic energy exchanges of locally
  confined hard spheres  \label{sec:Application}}

\paragraph{Stochastic kernel.}
In order to reduce the problem to the calculation of dimensionless
quantities at equilibrium, we set $\ell\equiv1$ and rescale the
stochastic kernel  according to 
\begin{equation} 
  W(\epsilon_{0},\epsilon_{1}\vert \epsilon_{0}-\eta,\epsilon_{1}+\eta)
  = \nu(T) \, \beta \, w(\beta\epsilon_{0},\beta\epsilon_{1}\vert
  \beta\epsilon_{0}-\beta\eta,\beta\epsilon_{1}+\beta\eta)
  \,,
  \label{eq:W-w} 
\end{equation} 
in terms of the inverse temperature $\beta=T^{-1}$ and
the corresponding equilibrium average of the binary collision
frequency~\eqref{eq:nu}, 
\begin{equation} 
  \nu(T) = \avgeq{ \nu(\epsilon_{0},\epsilon_{1}) } \equiv \sqrt{T}\, ,
\end{equation}
which now has the dimensions of the heat conductivity. Notice the
second equality can be assumed without loss of generality, under a
proper rescaling of time; see reference
\cite[Section~6]{Gaspard:2009P08020}.  

Introducing the dimensionless quantities $e_l \equiv \beta \,
\epsilon_{l}$ and $h \equiv \beta \, \eta$, the rescaled stochastic
kernel for a system of hard spheres ($D=3$) can be written out as  
\begin{multline} 
  w(e_{0},e_{1}\vert e_{0}-h,e_{1}+h) 
    =
    \frac{1}{(2\pi)^{3/2}\sqrt{e_{0}\,e_{1}}} \int_{\mathbf{n}_{01}\cdot
    \mathbf{u}_{01}>0} \rmd\mathbf{u}_0\, \rmd\mathbf{u}_1 \, 
    \mathbf{n}_{01}\cdot\mathbf{u}_{01}
    \, \deltad(e_{0}-\mathbf{u}_0^2) 
    \cr
    \times \deltad(e_{1}-\mathbf{u}_1^2) \,
    \deltad\left[h-(\mathbf{n}_{01}\cdot\mathbf{u}_0)^2+(\mathbf{n}_{01}
    \cdot\mathbf{u}_1)^2\right] , 
\end{multline} 
where $\mathbf{u}_{01}=\mathbf{u}_0-\mathbf{u}_1$ is
proportional to the relative velocity between the two colliding balls
and $\mathbf{n}_{01}$ is a three-dimensional unit vector joining the
centers of the balls $0$ and $1$.  The explicit form of the stochastic
kernel is given by 
\begin{equation} 
  \label{eq:kernel}
  w(e_{0},e_{1}\vert e_{0}-h,e_{1}+h)
  =\sqrt{\frac{\pi}{8}}\times
  \begin{cases}
    \sqrt{\frac{e_{1}+h}{e_{0}\,e_{1}}}\, , &
    -e_{1}<h<\mathrm{min}(0,e_{0}-e_{1}) \, , \\
    \frac{1}{\sqrt{\mathrm{max}(e_{0},e_{1})}}\, , & 
    \mathrm{min}(0,e_{0}-e_{1})<h<\mathrm{max}(0,e_{0}-e_{1}) \, , \\
    \sqrt{\frac{e_{0}-h}{e_{0}\,e_{1}}}\, , &
    \mathrm{max}(0,e_{0}-e_{1})<h<e_{0} \, ,
  \end{cases}  
\end{equation} 
and zero otherwise. Equations \eqref{eq:nu}-\eqref{eq:h} become
\begin{align}
  \label{eq:3dnu}
  \nu(e_{0},e_{1})  
  &= 
    \frac{\sqrt{2\pi}}{12}
    \frac{e_{0} + e_{1} + 2\,\mathrm{max}(e_{0}, e_{1})}
    {\mathrm{max}({e_{0}}, {e_{1}})^{1/2}}
    \,,
  \\
  \label{eq:3dh}
  h(e_{0},e_{1}) 
  &=
    \frac{\sqrt{2\pi}}{420}
    \frac{11 (e_{0}^{3} + e_{1}^{3}) + 
    7\, e_{0} \, e_{1}[ 3 (e_{0} + e_{1}) - 8\,\mathrm{max}(e_{0}, e_{1})]
    + 24\,\mathrm{max}(e_{0}, e_{1})^{3}
    } 
    {\mathrm{max}({e_{0}}, {e_{1}})^{1/2}}
    \,,
\end{align}
with the associated current \eqref{eq:loc_av_curr},
\begin{equation}
  \label{eq:3dcurrent}
  j(e_{0}, e_{1}) = 
  \tfrac{1}{2} (e_{0} - e_{1})
  \nu(e_{0},e_{1})  
  \,;
\end{equation}
see reference~\cite{Gaspard:2009P08020}. 

We note that the stochastic kernel is both symmetric with respect to
space inversion and time reversal (or detailed balance): 
\begin{equation} 
  \label{eq:wsym}
  \begin{split}
  & w(e_{0},e_{1}\vert e_{0}-h,e_{1}+h)
    =w(e_{1},e_{0}\vert e_{1}+h,e_{0}-h) \,,
    \\ 
  & \sqrt{e_{0}\,e_{1}} \, w(e_{0},e_{1}\vert e_{0}-h,e_{1}+h)
    =\sqrt{(e_{0}-h)(e_{1}+h)} \, w(e_{0}-h,e_{1}+h\vert e_{0},e_{1}) \, . 
  \end{split}
\end{equation} 

\paragraph{Implementation of the variational formula.}

In terms of the dimensionless stochastic kernel introduced in
equation~\eqref{eq:W-w}, the variational formula~\eqref{eq:formula}
reads   
\begin{equation}
  \label{eq:kappa_var} 
  \kappa(T) = \tfrac{1}{2} \sqrt{T}\, \mathrm{inf}_f 
  \avgeq{ \int dh \,
    w(e_{0},e_{1}\vert e_{0}-h,e_{1}+h)\Big(h + 
    \sum_{k\in\mathbb{Z}} D_{0,1,h} \htau^{k} f
    \Big)^2 } .
\end{equation}

Trial functions $f$ can be expanded in terms of the generalized
Laguerre polynomials $\La{n}(x)$, which form an orthogonal basis:
\begin{equation} 
  \int_0^{\infty} dx \, \sqrt{x} \, \rme^{-x} \, \La{m}(x) \,
  \La{n}(x) = \frac{1}{n!} \, \Gamma(n+3/2) \, \delta_{m,n} 
  \,,
\end{equation}
where $\delta_{m, n}$ denotes the Kronecker symbol and  the Gamma
function of half-integer argument can be expressed as 
\begin{equation} 
  \Gamma(n+3/2) = \sqrt{\pi} \, \frac{(2n+1)!!}{2^{n+1}} 
  \,.
\end{equation} 
It is convenient to define a basis of orthonormal functions\footnote{
  The first few such polynomials are given by 
  \begin{align*} 
    J_0(x) &= 1 \, , 
    &
      J_1(x) &= \frac{1}{\sqrt{6}} (3-2x)\, , \\     
    J_2(x) &= \frac{1}{2\sqrt{30}} (15-20x + 4x^2)\, , 
    &
      J_3(x) &= \frac{1}{12\sqrt{35}} (105-210x + 84x^2-8x^3)\, ,\dots 
  \end{align*}
}
\begin{equation} 
  J_{n}(x) = \sqrt{\frac{2^n\, n!}{(2n+1)!!}} \, \La{n}(x) \, , 
\end{equation} 
which satisfy the
orthonormality condition 
\begin{equation} 
  \label{eq:orthoJnJm}
  \frac{2}{\sqrt\pi}\int_0^{\infty} dx \, \sqrt{x} \rme^{-x} \, J_m(x)
  \, J_n(x) = \delta_{m,n} \,, 
\end{equation} 
with weight function given by a Gamma distribution of shape
parameter $\tfrac{3}{2}$ and scale parameter unity (which coincides
with the single-cell marginal of the equilibrium distribution
\eqref{eq:P_eq} at unit temperature). 

To find the infimum of the variational formula \eqref{eq:kappa_var},
trial functions with an increasing number $r$ of variables can be
successively considered: 
\begin{align}
  f(e_{1},\dots,e_r) 
  = \sum_{n_1, \dots, n_r}
  \gammars{r,\,\infty}_{n_1, \dots, n_r} \, J_{n_1}(e_{1}) \dots J_{n_r}(e_r) \, .
  \label{eq:fr}
\end{align} 
We refer to the number of variables $r$ as the order of the
approximation. Functions of order $r=2$ are obviously a subset of
functions of order $r\geq3$ and similarly for every $r$. 

The $\infty$ superscript on the coefficients appearing in equation
\eqref{eq:fr} refers to the unrestricted sum over
$n_{i}\in\mathbb{N}$, $i=1,\dots,r$. That is, for a fixed order $r$,
the sum in equation~\eqref{eq:fr} runs over orthonormal Laguerre
polynomials of all degrees. By restricting this 
sum to a maximum degree $s$, such that $n_{1}+\dots+n_{r}\leq s$, we
obtain a function space of finite dimension for which the variational
formula boils down to computing the infimum of a quadratic form in the
coefficients $\gamma^{(r,\,s)}_{n_1,\dots,n_{r}}$. Its solution thus
yields an approximation, $\kappad^{(r,\,s)}$, to the actual dynamical
contribution to the heat conductivity in the form of an upper bound,
\begin{equation}
  \label{eq:kappadbound}
  \kappad < \kappad^{(r,\,s)}\,,
\end{equation}
obtained by restricting the computation of the infimum in
\eqref{eq:kappa_var} to the set of multivariate polynomials of $r$
variables and degree $s$.

For every order $r$, the optimal upper bound, $\kappad^{(r,\,\infty)}$, is 
obtained by letting $s \to \infty$. The convergence to the actual
infimum is subsequently obtained by letting $r\to\infty$. The
dynamical contribution to the heat conductivity is thus estimated as
the result of a double extrapolation, first in the degree $s$, then in
the order $r$.

Moreover, the location of the infimum in the infinite-dimensional
space of trial functions allows us to compute the part of the steady
state that contributes to the current, equation \eqref{eq:stst2},
whose two-cell marginal  enters the expression \eqref{eq:kappa-stst}
of the heat conductivity. We can therefore retrieve the dynamical
contribution to the heat conductivity by integrating the current: 
\begin{align}
  \label{eq:kappa-ststexplicit}
  \frac{\kappad(T)}{\sqrt{T}} 
  &
  = - \sum_{n_{0}, n_{1}}
  \gamma_{\dots,0,n_{0},n_{1},0,\dots}
  \avgeq{j(e_{0}, e_{1}) 
    J_{n_{0}}(e_{0}) J_{n_{1}}(e_{1})
    }
    \,,\cr
    &
    =
      - \frac{1}{12\sqrt{5}} \gamma_{\dots,0,2,1,0,\dots}
      - \frac{1}{4\sqrt{210}} \gamma_{\dots,0,3,1,0,\dots}
      - \frac{1}{32\sqrt{42}} \gamma_{\dots,0,3,2,0,\dots}
      \cr
  &\quad
    - \frac{\sqrt{5}}{64\sqrt{21}} \gamma_{\dots,0,4,1,0,\dots}
    - \frac{\sqrt{7}}{384\sqrt{3}} \gamma_{\dots,0,4,2,0,\dots}
    - \frac{1}{512\sqrt{2}} \gamma_{\dots,0,4,3,0,\dots}
      - \dots
\end{align}
The correspondence between the coefficients of two non-trivial indices
$\gamma_{\dots,0, n_{0}, n_{1}, 0, \dots}$ which enter this expression and
the finite $r$ and $s$ coefficients which realize the infimum
\eqref{eq:kappa_var} over multivariate polynomials of $r$ variables
and degree $s$ is as follows
\begin{align}
  \label{eq:gammars}
  \gamma_{\dots,0, n_{0}, n_{1}, 0,  \dots} 
  & = \lim_{r,\,s\to\infty}
    \left( \gamma^{(r,\,s)}_{n_{0}, n_{1}, 0,  \dots, 0} + \dots +
    \gamma^{(r,\,s)}_{0,  \dots, 0, n_{0}, n_{1}} \right)
    \,,
    \cr
  &
    = \lim_{r,\,s\to\infty} (r-1) \gamma^{(r,\,s)}_{n_{0}, n_{1}, 0,  \dots, 0}\,,
    \cr
  &
    \equiv \lim_{r,\,s\to\infty} \gamma^{(r,\,s)}_{n_{0}, n_{1}}\,,
\end{align}
where the second line follows by identity of the $r-1$
coefficients $\gamma^{(r,\,s)}_{n_{0}, n_{1}, 0,  \dots, 0} = \dots =
\gamma^{(r,\,s)}_{0,  \dots, 0, n_{0}, n_{1}}$, and the third line
introduces a convenient shorthand notation.

In subsection \ref{sec:Calculation-(r=2)}, we shall begin by detailing
the first steps of the calculation for $r=2$, starting
from $s=2, 3$ and $4$ (as explained above, trial functions with $r=1$
should not be considered because they do not modify the value of the
conductivity). This provides the basis for a systematic extension up
to $s=15$, which we subsequently extrapolate to $s\to\infty$, owing to
their fast convergence. We then go on in subsection
\ref{sec:Calculation-(r>2)} to work out the systematic extension of
these results to multivariate test functions and so obtain an estimate
of $\kappad = \lim_{r,\,s\to\infty} \kappad^{(r,\,s)}$. 

\subsection{Restriction to bivariate trial functions ($r=2$)}
\label{sec:Calculation-(r=2)}

Let us begin by restricting the computation of the infimum in
equation~\eqref{eq:kappa_var} to trial functions $f$ depending on two
variables only. For such trial functions, we have that 
\begin{multline} 
  \sum_{k\in\mathbb{Z}} D_{0,1,h} \htau^{k} f = 
  f(e_{-1},e_{0}-h) +
  f(e_{0}-h,e_{1}+h)+f(e_{1}+h,e_2) \cr
  - f(e_{-1},e_{0}) - f(e_{0},e_{1}) -f(e_{1},e_2)
  \, .  
\end{multline} 
The function $f$ is expanded according to equation~\eqref{eq:fr} with
$r=2$ so that the variational formula~\eqref{eq:kappa_var} becomes a
quadratic form in the coefficients $\gammars{2,\,s}_{m,n}$: 
\begin{multline} 
  \label{eq:kappa_var_r2} 
  \frac{\kappad^{(2,\,s)}(T)}{\sqrt{T}} 
  = \mathrm{inf}_{\{\gammars{2,\,s}_{m,n}\}}
    \Bigg[ \sqrt{\frac{3}{2}} \sum_{\substack{m,n = 0\\m+n\leq s}}^{s}
    \gammars{2,\,s}_{m,n} \, ( \delta_{m,0} \, A_{n,0,1,0} + A_{m,n,1,0} + 
    \delta_{n,0} \, A_{0,m,1,0} ) 
    \cr
    +\frac{1}{2} 
    \!\!\!\!
    \sum_{\substack{m,n,p,q = 0\\m+n\,\&\,p+q\leq s}}^{s}
    \!\!\!\!
    \gammars{2,\,s}_{m,n} \, \gammars{2,\,s}_{p,q} \, ( \delta_{m,p} \,
    A_{n,0,q,0} + \delta_{m,0} \, A_{n,0,p,q} + \delta_{m,0} \,
    \delta_{q,0} \, A_{n,0,0,p} 
    + \delta_{p,0} \, A_{m,n,q,0} 
    \cr 
    + A_{m,n,p,q} + \delta_{q,0} \, A_{m,n,0,p} 
    + \delta_{n,0} \, \delta_{p,0} \,
     A_{0,m,q,0} + \delta_{n,0} \, A_{0,m,p,q} + \delta_{n,q} \,
    A_{0,m,0,p} )\Bigg]\,, 
\end{multline}
which is obtained by substituting $h = \sqrt{{3}/{2}} \left[
  J_1(e-h)-J_1(e)\right]$ for the terms linear in $h$ and, for $i=-1$
and $i=2$, using the orthonormality of the polynomial basis 
\eqref{eq:orthoJnJm},
\begin{equation} 
  \begin{split}
  \langle J_m(e_i)\rangle_\mathrm{eq} &= \delta_{m,0} \, ,
  \\ 
  \langle J_m(e_i)J_n(e_i)\rangle_\mathrm{eq} &= \delta_{m,,n} \, .
  \end{split}
\end{equation}

The coefficients $A_{m,n,p,q}$ which enter
equation~\eqref{eq:kappa_var_r2} are numbers, defined by 
\begin{multline} 
  A_{m,n,p,q} \equiv
  \frac{4}{\pi}
  \int_{0}^{\infty} \rmd e_{0} \, \rmd e_{1} 
  \int_{-e_{1}}^{e_{0}} \rmd h \, \sqrt{e_{0}\,e_{1}}\,\rme^{-e_{0}-e_{1}}\, 
  w(e_{0},e_{1}\vert e_{0}-h, e_{1}+h) \cr
  \times \big[ J_m(e_{0}-h)J_n(e_{1}+h)-J_m(e_{0})J_n(e_{1})\big] 
  \big[ J_p(e_{0}-h)J_q(e_{1}+h)-J_p(e_{0})J_q(e_{1})\big] .
  \label{eq:A} 
\end{multline} 
They obey the symmetry relations,
\begin{equation} 
  \label{eq:Asymmetries} 
  \begin{split}
    &A_{m,n,p,q}=A_{n,m,q,p} \, , \\
    &A_{m,n,p,q}=A_{p,q,m,n} \, ,
  \end{split}
\end{equation} 
which are a consequence of the symmetries \eqref{eq:wsym} of the
stochastic kernel. Equation \eqref{eq:A} also immediately implies 
\begin{equation}
  \label{eq:A00} 
  A_{m,n,0,0}=0\,.
\end{equation}

Some particular non-trivial values of the coefficients~\eqref{eq:A} are given by 
\begin{equation} 
  \label{eq:A11} 
  \begin{aligned}
  A_{1,0,1,0} &= -A_{1,0,0,1} = \frac{4}{3} \, , 
  &\qquad
  A_{1,1,1,1} &= \frac{16}{9}\,,
  \\
  A_{2,0,1,0} &= - A_{2,0,0,1} = -\frac{2}{3\sqrt5} \, , 
  &\qquad
  A_{2,0,2,0} &= \frac{31}{15} \, , 
  \\
  A_{2,0,2,0} &= -1\, , 
  &\qquad
  A_{2,0,1,1} &= -\frac{16}{3\sqrt30} \, ,
  \end{aligned}
\end{equation}
from which all coefficients $A_{m,n,p,q}$ such that $m + n$ and $p + q
\leq 2$ can be obtained using equations \eqref{eq:Asymmetries} and
\eqref{eq:A00}.

For general indices, the coefficients can be conveniently expressed as
follows: 
\begin{multline}
  \label{eq:Aexplicit}
  A_{m,n,p,q} = 
  16 \left[ \frac{2^{(m + n + p + q)} m\,!\,n\,!\,p\,!\,q\,!}
    {(2m+1)!!\,(2n+1)!!\,(2p+1)!!\,(2q+1)!!} \right]^{1/2}
  \cr
  \times
  \sum_{i = 0}^{m} \sum_{j = 0}^{n} \sum_{k = 0}^{p} \sum_{l = 0}^{q}
  \frac{2^{i+j+k+l}}
  {i!\,j!\,k!\,l!}  
  \binom{m + \tfrac{1}{2}}{m - i}
  \binom{n + \tfrac{1}{2}}{n - j}
  \binom{p + \tfrac{1}{2}}{p - k}
  \binom{q + \tfrac{1}{2}}{q - l}
  \cr
  \times
  \left[
    \partial_{\beta_{1}}^{i+k} 
    \partial_{\beta_{2}}^{j+l}
    -
    \partial_{\beta_{1}}^{i} 
    \partial_{\beta_{2}}^{j}
    \partial_{\beta_{3}}^{k} 
    \partial_{\beta_{4}}^{l}
  \right]
  \left.
  \frac{\sqrt{\beta_{1} + \beta_{2} +\beta_{3} +\beta_{4}}}
  {(\beta_{1} + \beta_{3}) (\beta_{1} + \beta_{4}) 
    (\beta_{2} +\beta_{3}) (\beta_{2} + \beta_{4})}
  \right|_{\beta_{1} = \dots = \beta_{4} = 1}
  \,,
\end{multline}
which allows for a fast tabulation (for $m+n$ and $p+q\leq s$ and $s
\leq 15$ which is the largest degree we work with).

Moreover, the identity between generalized Laguerre functions,
\begin{equation} 
  \sum_{m=0}^n L_m^{(\alpha)}(x-z) \,
  L_{n-m}^{(\beta)}(y+z) = \sum_{m=0}^n L_m^{(\alpha)}(x) \,
  L_{n-m}^{(\beta)}(y) 
  \,,
\end{equation} 
which follows from a special case of an addition formula \cite[p. 192,
equation~(41)]{Bateman:1953wkb}, 
implies the following sum rule: 
\begin{equation} 
  \sum_{m=0}^n A_{m,n-m,p,q} = \sum_{p=0}^q A_{m,n,p,q-p} = 0
  \,.
\end{equation} 
Thus, in particular,
\begin{equation} 
  A_{0,1,p,q} = A_{1,0,q,p} = -A_{1,0,p,q} = -A_{0,1,q,p} \, .  
\end{equation}

\paragraph{Calculation for degree up to $s=2$.}

In this case all coefficients are in fact trivial, so that no non-trivial
contribution to the heat conductivity arises at this degree. 
Indeed, the coefficients $\gammars{2,2}_{0,i}$ and
$\gammars{2,2}_{i,0}$ ($i=1,2$) need not be considered because they
enter into the expansion of a univariate function, which can always be
eliminated; see equation~\eqref{eq:nosinglevarfunction}. The last
remaining coefficient is $\gammars{2,2}_{1,1}$, which would bring about
a symmetric contribution to the infimum and can therefore be
also eliminated.

These observations are verified by explicit calculation of equation
\eqref{eq:kappa_var_r2}\footnote{We henceforth set $T\equiv1$ and omit the
  explicit temperature dependence.},  
\begin{equation} 
  \label{eq:r2s2kappad}
  \kappad^{(2,2)} = \mathrm{inf}_{\gammars{2,2}_{1,1}} 
  \bigg[
  \frac{20}{9}{\gammars{2,2}_{1,1}}^2
  -\frac{16\sqrt{3}}{45}\gammars{2,2}_{1,1}(\gammars{2,2}_{0,2} +
  \gammars{2,2}_{2,0}) 
  + \frac{16}{15}(\gammars{2,2}_{0,2} + \gammars{2,2}_{2,0})^2
  \bigg],
\end{equation} 
which is trivial and attained when
\begin{equation}
  \label{eq:r2s2gamma}
  \begin{split}
      \gammars{2,2}_{1,1} &= 0\,,\\
      \gammars{2,2}_{2,0} &= -\gammars{2,2}_{0,2}\,.
  \end{split}
\end{equation}

\paragraph{Calculation for degree up to $s=3$.}

There exist a priori two non-trivial coefficients
$\gammars{2,3}_{m,n}$ contributing to equation~\eqref{eq:kappa_var_r2}, 
i.e.~$\gammars{2,3}_{1,2}$ and $\gammars{2,3}_{2,1}$. They must
however be opposite to one another by antisymmetry of the function
realizing the infimum, 
\begin{equation}
  \label{eq:r2s3asym}
    \gammars{2,3}_{1,2} = -\gammars{2,3}_{2,1}
    \,.
\end{equation}
Equation~\eqref{eq:kappa_var_r2} can therefore be simplified to 
\begin{align} 
  \label{eq:r2s3kappad}
  \kappad^{(2,3)}
  & =\mathrm{inf}_{\gammars{2,3}_{2,1}}
  \bigg(
    -\frac{1}{6\sqrt{5}}\,\gammars{2,3}_{2,1} + 
    \frac{335}{48} \,{\gammars{2,3}_{2,1}}^2
  \bigg) , 
    \cr
  &
    = - \frac{1}{5025} = - 0.000\,199\,005 ,
\end{align} 
which is attained when
\begin{equation} 
  \label{eq:r2s3gamma}
  \gammars{2,3}_{2,1} = \frac{4}{335\sqrt{5}} \, .
\end{equation} 

Equation~\eqref{eq:r2s3kappad} presents the crudest approximation to
the heat conductivity beyond the sole static contribution. It provides
an upper bound on the value of the conductivity and already confirms
that the dynamical contribution is not trivial. To lower this bound
and improve it, we must increase the degree or the order of the trial
functions. We start by considering the former possibility; the latter will
be addressed in \sref{sec:Calculation-(r>2)}.

\paragraph{Calculation for degree up to $s=4$.}

There are two distinct coefficients $\gammars{2,4}_{m,n}$ contributing to
equation~\eqref{eq:kappa_var_r2} when the degrees of the trial
polynomials are restricted to $m+n\leq s=4$: $\gammars{2,4}_{2,1} \ (= 
-\gammars{2,4}_{1,2})$ and $\gammars{2,4}_{3,1}\
(=\gammars{2,4}_{1,3})$. The second approximation to the variational
formula is thus given by  
\begin{align} 
  \label{eq:r2s4kappad}
  \kappad^{(2,4)}
  &
    = \mathrm{inf}_{\{\gammars{2,4}_{m,n}\}} \bigg( 
    - \frac{1}{6\sqrt{5}}\gammars{2,4}_{2,1}
    -\frac{1}{2\sqrt{210}}\gammars{2,4}_{3,1} 
    + \frac{335}{48} {\gammars{2,4}_{2,1}}^2 
    \cr
  &
    \hspace{4cm}
    - \frac{121}{8\sqrt{42}} \gammars{2,4}_{2,1}\,\gammars{2,4}_{3,1} 
    + \frac{2621}{336}{\gammars{2,4}_{3,1}}^2 \bigg) \,,
    \cr
  &
    = -\frac{367}{1\,351\,695} \simeq - 0.000\,271\,511 \,,
\end{align}
which is reached for the coefficients
\begin{equation}
  \label{eq:r2s4gamma}
  \begin{split}
    \gammars{2,4}_{2,1} =
    & \frac{236\sqrt{5}}{90\,113} \, , \\ 
    \gammars{2,4}_{3,1} 
    =&
    \frac{96}{90\,113}\sqrt{\frac{42}{5}} \, .
  \end{split}
\end{equation} 

Comparing equations \eqref{eq:r2s3kappad} and \eqref{eq:r2s4kappad}, 
we observe that, as expected, the latter approximation to the
dynamical contribution to the heat conductivity is lower (and
substantially so) than the former. Likewise, the value of
$\gammars{2,4}_{2.1}$ in equation~\eqref{eq:r2s4gamma} is larger
(although only by about 10\%) than the $s=3$ value,
equation~\eqref{eq:r2s3gamma}.

\paragraph{Calculations for higher degree values.}

It is in principle straightforward to extend the computation described
above to higher degrees $s$. One is however limited by the rapid
growth of the number of coefficients $A_{m,n,p,q}$ which must be
computed to write out the corresponding approximation to equation
\eqref{eq:kappa_var_r2}. Here we limit our investigations to degrees
$s\leq15$. 

\begin{table}[hbt]
  \begin{center}
    \begin{tabular}{|c|c|@{}c@{}|@{}c@{}|@{}c@{}|@{}c@{}|@{}c@{}|@{}c@{}|} 
      \hline
      $s$ & $-10^{4} \kappad^{(2,\,s)}$ & $10^{3} \gammars{2,\,s}_{2,1}$ &
      $10^{3} \gammars{2,\,s}_{3,1}$ & $10^{3} \gammars{2,\,s}_{4,1}$ & 
      $10^{3} \gammars{2,\,s}_{3,2}$ & $10^{3} \gammars{2,\,s}_{5,1}$ & 
      $10^{3} \gammars{2,\,s}_{4,2}$ \\
      \hline
      $3$ & $1.990\,05$  & $5.339\,86$  &  & & & & \\
      $4$ & $2.715\,11$  & $5.831\,30$  & $2.087\,62$  &  & & & \\
      $5$ & $2.985\,04$  & $5.957\,36$  & $3.487\,94$  & $1.601\,55$ & 
      $0.850\,87$ & & \\
      $6$ & $3.095\,00$  & $5.984\,59$ & $3.582\,45$ & $1.853\,21$ & 
      $0.981\,92$ & $0.828\,39$ & $0.751\,69$ \\
      $7$ & $3.143\,97$  & $5.993\,60$ & $3.611\,28$ & $1.921\,29$ & 
      $1.016\,55$ & $0.978\,29$  & $0.883\,95$ \\
      $8$ & $3.167\,57$  & $5.997\,08$ & $3.621\,73$ & $1.944\,26$ & 
      $1.027\,95$ & $1.023\,45$  & $0.922\,57$ \\
      $9$ & $3.179\,74$  & $5.998\,59$ & $3.626\,04$ & $1.953\,23$ & 
      $1.032\,30$ & $1.040\,00 $  & $0.936\,29$ \\
      $10$ & $3.186\,40$ & $5.999\,31$ & $3.628\,00$ & $1.957\,16$ & 
      $1.0341\,7$ & $1.046\,91$  & $0.941\,85$ \\
      $11$ & $3.190\,22$ & $5.999\,68$ & $3.628\,98$ & $1.959\,04$ & 
      $1.035\,04$ & $1.050\,09$  & $0.944\,34$ \\
      $12$ & $3.192\,52$ & $5.999\,89$ & $3.629\,50$ & $1.960\,01$ & 
      $1.035\,48$ & $1.051\,68$  & $0.945\,55$ \\
      $13$ & $3.193\,96$ & $6.000\,01$ & $3.629\,80$ & $1.960\,54$ & 
      $1.035\,72$ & $1.052\,53$ & $0.946\,19$ \\
      $14$ & $3.194\,88$ & $6.000\,08$ & $3.629\,97$ & $1.960\,85$ & 
      $1.035\,86$ & $1.053\,01$ & $0.946\,54$ \\
      $15$ & $3.195\,49$ & $6.000\,12$ & $3.630\,08$ & $1.961\,03$ & 
      $1.035\,94$ & $1.053\,29$ & $0.946\,75$ \\
      \hline
      $\infty$ & $3.197\,13(5)$  & $6.000\,30(8)$  & $3.630\,26(1)$  & 
      $1.961\,26(7)$ & $1.036\,04(3)$  & $1.053\,6(1)$  & $0.946\,96(8)$ \\
      \hline
    \end{tabular}
  \end{center}
  \caption{Decimal approximations of the upper bound $\kappad^{(2,\,s)}$
    of the dynamic contribution to the conductivity (multiplied by
    -10\,000) and of the first few non-trivial coefficients $\gammars{2,\,s}_{m,n}$
    (multiplied by 1000) contributing to approximations of the
    infimum in equation~\eqref{eq:kappa_var_r2} by bivariate polynomials
    of degrees $s=3,\dots,\,15$. The last line reports the estimated
    asymptotic values obtained from a nonlinear regression model; see
    details in the text. The digits in brackets indicate the estimated
    uncertainty on the last reported digit \cite{Taylor:1999nist}.
  }  
  \label{tab:gammar2}
\end{table}

\begin{figure}[htb]
  \centering
  \begin{subfigure}[t]{0.495\linewidth}
    \includegraphics[width=\textwidth]{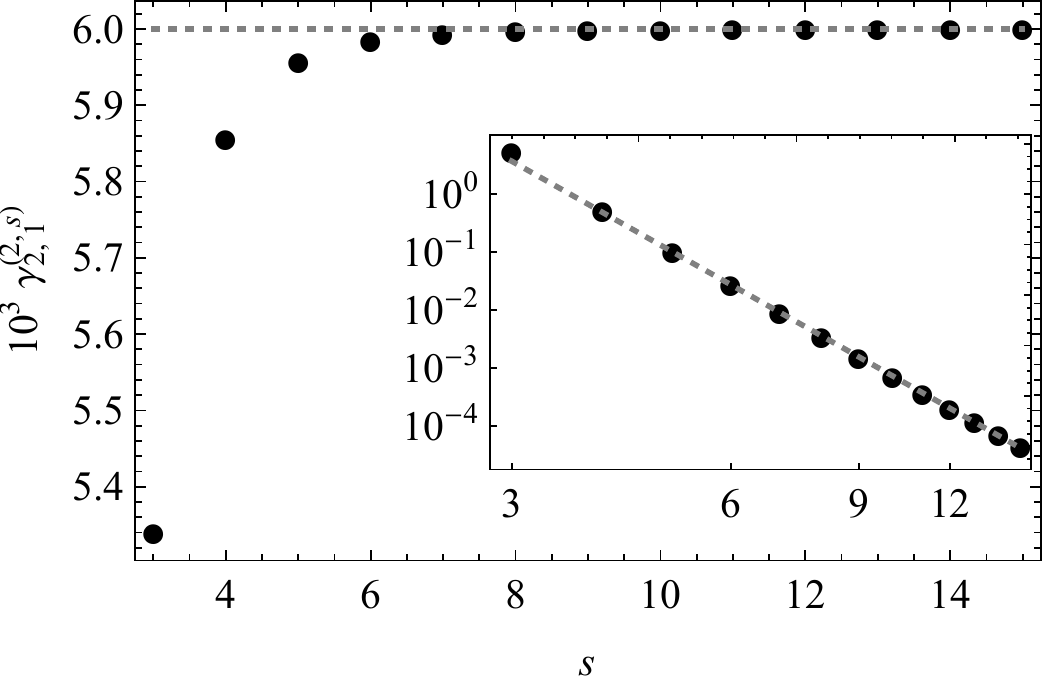}
    \caption{$\gammars{2,\,s}_{2,1}$ v. $s$}
    \label{fig:gammar2_m2n1}
  \end{subfigure}
  \begin{subfigure}[t]{0.495\linewidth}
    \includegraphics[width=\textwidth]{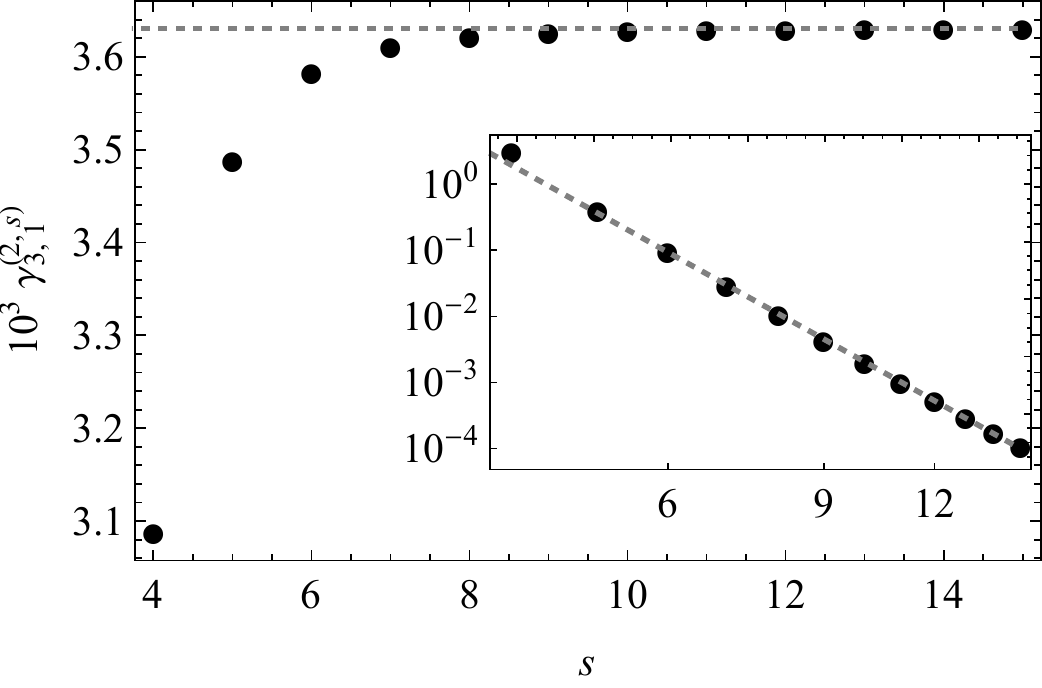}
    \caption{$\gammars{2,\,s}_{3,1}$ v. $s$}
    \label{fig:gammar2_m3n1}
  \end{subfigure}
  \begin{subfigure}[t]{0.495\linewidth}
    \includegraphics[width=\textwidth]{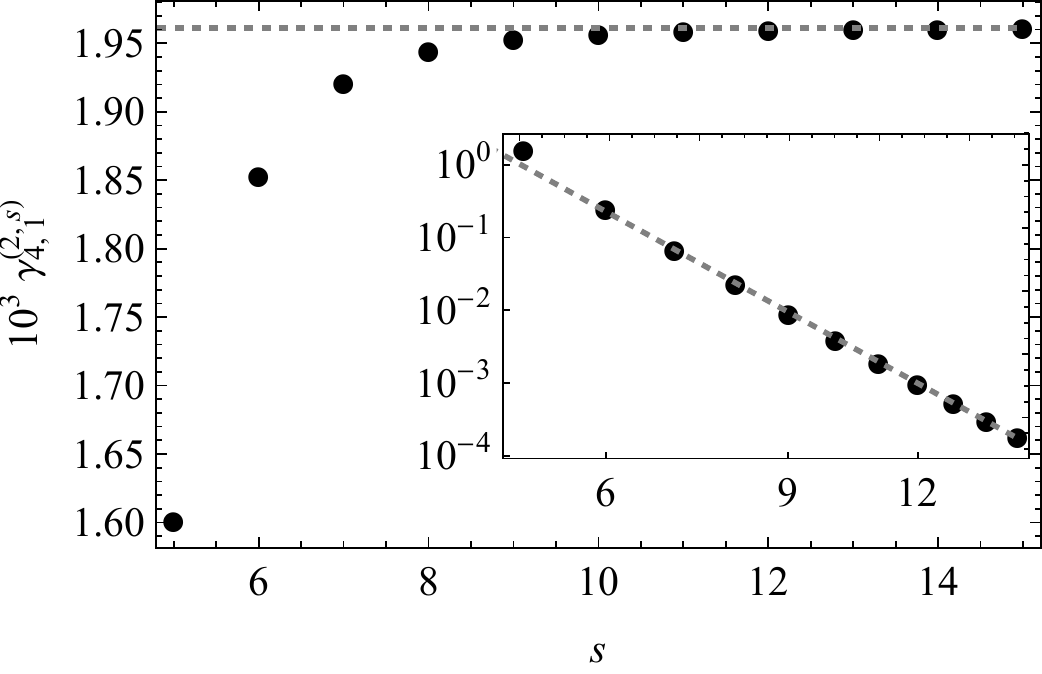}
    \caption{$\gammars{2,\,s}_{4,1}$ v. $s$}
    \label{fig:gammar2_m4n1}
  \end{subfigure}
  \begin{subfigure}[t]{0.495\linewidth}
    \includegraphics[width=\textwidth]{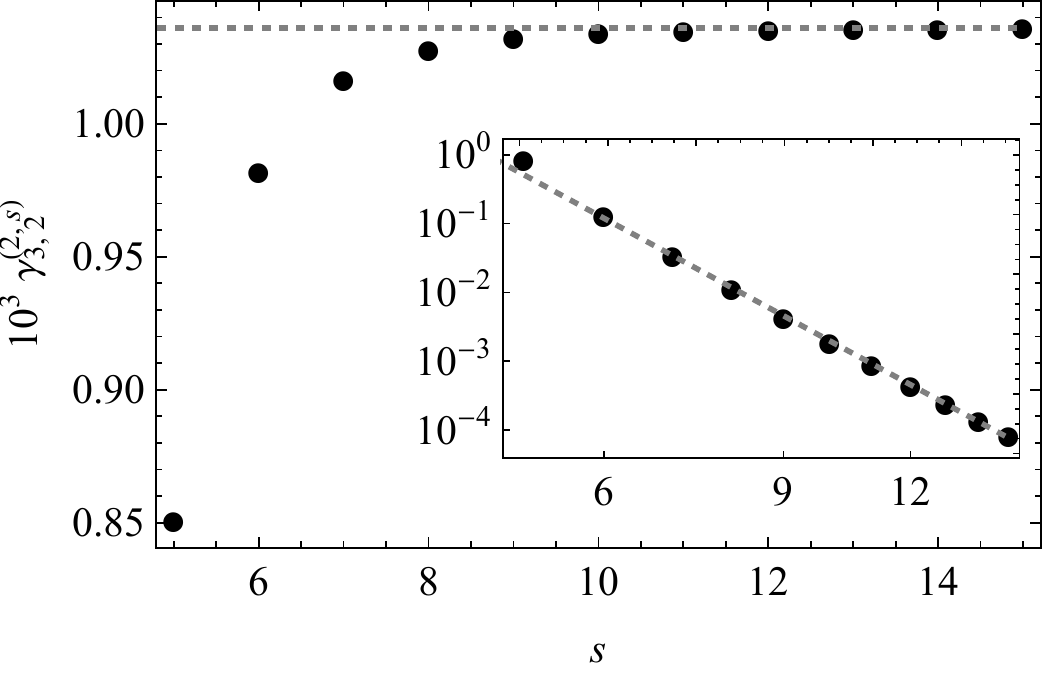}
    \caption{$\gammars{2,\,s}_{3,2}$ v. $s$}
    \label{fig:gammar2_m3n2}
  \end{subfigure}
  \caption{Graphical representations of the behaviour of the coefficients
    $\gammars{2,\,s}_{m,n}$ (such that $m+n\leq 5$) as $s$ increases. The dotted
    straight lines indicate the results of the $s\to\infty$ extrapolations reported
    in table~\ref{tab:gammar2}. The insets show on a log-log scale
    the decay of the increments $\gammars{2,\,s+1}_{m,n} -
    \gammars{2,\,s}_{m,n}$ as $s$ increases.
  } 
  \label{fig:fit_gammar2}
\end{figure}

In table~\ref{tab:gammar2}, we list, in the second column, the values
of the upper bounds on the dynamic contribution to the heat 
conductivity $\kappad^{(2,\,s)}$ when restricting its computation to
bivariate polynomials of degrees $s = 3,\dots,\,15$. The
remaining columns show the values of some of the non-trivial 
coefficients $\gammars{2,\,s}_{m,n}$ for which the infimum is
reached (the list of coefficients is limited to indices $m$ and $n$
such that $m+n\leq 6$). The values of the upper bounds on the dynamic
contribution to the heat conductivity thus coincide with the values 
inferred from the transposition of equation
\eqref{eq:kappa-ststexplicit} to coefficients of finite order $r=2$ and
corresponding degree $s$.

The last line of table~\ref{tab:gammar2} lists the results of the
infinite-degree extrapolations of $\kappad^{(2,\,s)}$ and
$\gammars{2,\,s}_{m,n}$ with error estimates on their last digits. To
obtain those estimates, we first consider the coefficients
$\gammars{2,\,s}_{m,n}$ and notice their rapid convergence to
asymptotic values as $s$ increases. To evaluate this convergence, we
plot in \fref{fig:fit_gammar2} the graphs of the first four
coefficients as functions of the degree $s$ (similar results are
obtained for the other coefficients). The insets of these figures exhibit the
decay of the increments $\gammars{2,\,s+1}_{m,n} - \gammars{2,\,s}_{m,n}$ as
$s$ increases, which appear to follow simple power laws of the form
$b \, s^{-c}$, and are readily fitted by linear regressions (we
remove the values $s=m+n$ from the fitted sequences). The
results give the respective exponent values 
\[
  c = 
  \begin{cases}
    7.07 \pm 0.03 \quad \mbox{(\fref{fig:gammar2_m2n1})},      \\
    7.47  \pm 0.04 \quad \mbox{(\fref{fig:gammar2_m3n1})},      \\
    7.85 \pm 0.06 \quad \mbox{(\fref{fig:gammar2_m4n1})},      \\
    8.03 \pm 0.06 \quad \mbox{(\fref{fig:gammar2_m3n2})}.
  \end{cases}
\]
The corresponding coefficients are respectively found to be $\log b =
9.07 \pm 0.06$,  $11.0 \pm 0.1$, $12.6 \pm 0.1$ and $12.3 \pm 0.1$. 
For each pair of parameters $b$ and $c$ thus obtained, we estimate
the $s\to\infty$ extrapolation of $\gammars{2,\,s}_{m,n}$ by adding to
the last computed value, $\gammars{2,15}_{m,n}$, the sum of the
modeled increments,  
\begin{equation}
  \label{eq:gammar2_modelincrements}
  \gammars{2,\infty}_{m,n} \approx \gammars{2,15}_{m,n}  + b \sum_{s =
    0}^\infty (s + 16)^{-c}
  \,.
\end{equation}
The results are:
\begin{alignat}{3}
  \label{eq:gammar2_modelincrementsresults}
  \gammars{2,\infty}_{2,1} &= 6.000\,21\,, 
  & \quad
  \gammars{2,\infty}_{3,1} &= 3.630\,26 \,,
  & \quad
  \gammars{2,\infty}_{4,1} &= 1.961\,33\,,
  \cr
  \gammars{2,\infty}_{3,2} &= 1.036\,07\,,
  & \quad
  \gammars{2,\infty}_{5,1} &= 1.053\,72\,,
  & \quad
  \gammars{2,\infty}_{4,2} &= 0.947\,04\,.
\end{alignat}

An alternative way of obtaining $s\to\infty$ extrapolations, which
avoids resorting to the increments is to model the coefficients
according to the power law $\gammars{r,\,s}_{m,n}  
\approx a - \rme^{\log b} \, s^{-c}$. The corresponding coefficients
(treating $\log b$ as such) can be
evaluated through nonlinear regressions\footnote{The results of
  nonlinear regressions were obtained using the statistical model
  analysis in Mathematica (\url{http://www.wolfram.com}). } of the
computed values.  This procedure yields the approximations: 
\begin{alignat}{2}
  \label{eq:gammar2_model}
  \gammars{2,s}_{2,1} & \approx 
  6.000\,30 - 274.409\, s^{-5.446\,98}\,,
  & \quad
  \gammars{2,s}_{3,1} & \approx 
  3.630\,26 - 2\,168.14\, s^{-5.984\,24}\,,
  \cr
  \gammars{2,s}_{4,1} & \approx 
  1.961\,26 - 11\,010.0\,s^{-6.436\,06}\,,
  & \quad
  \gammars{2,s}_{3,2} & \approx 
  1.036\,04 - 7\,585.76\,s^{-6.614\,03}\,,
  \cr
  \gammars{2,s}_{5,1} & \approx 
  1.053\,62 - 43\,746.1\,s^{-6.820\,7}\,,
  & \quad
  \gammars{2,s}_{4,2} & \approx 
  0.946\,96 - 60\,580.1\,s^{-7.079\,66}\,.
\end{alignat}
The constant terms are the values reported in
table~\ref{tab:gammar2}. Although both linear and nonlinear
regressions yield error estimates of the coefficients, we note that
such error estimates tend to underestimate the error; systematic
errors due to the chosen model must also be assessed. We thus prefer
using error estimates obtained by comparing the asymptotic values
given by the two different methods
\eqref{eq:gammar2_modelincrementsresults} and
\eqref{eq:gammar2_model}. The differences between the two are the
numbers in brackets reported in table~\ref{tab:gammar2}. 

\setcounter{footnote}{0}

\begin{figure}[htb]
  \centering
  \includegraphics[width=0.55\textwidth]{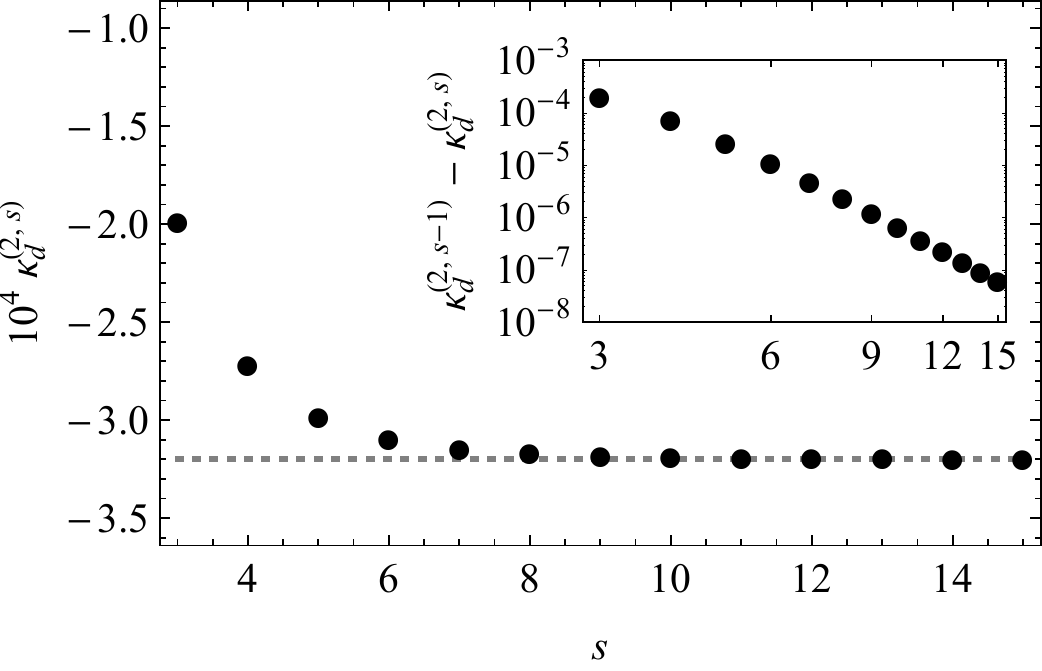}
  \caption{Graphical representations of the upper bounds
    $\kappad^{(2,\,s)}$ on the dynamical contribution to the heat
    conductivity restricted to bivariate polynomials of degree $s$ for
    $s=3,\dots,\,15$. The dotted straight line indicates the result of the
    $s\to\infty$ extrapolation reported in
    table~\ref{tab:gammar2}. The inset shows on a log-log scale the
    decay of the decrements $\kappad^{(2,\,s-1)} -\kappad^{(2,\,s)}$
    as $s$ increases.  
  } 
  \label{fig:kappadr2}
\end{figure}

Turning to $\kappad^{(2,\,s)}$, \fref{fig:kappadr2}, we do not
expect a simple power-law form will faithfully model this
quantity. Indeed, the exponent values reported on the right-hand side 
of equation~\eqref{eq:gammar2_model} differ significantly from each
other so that $\kappad^{(2,\,s)}$ should rather be thought of as a sum
of power laws with possibly many different exponents. Neither can we
rely on our limited estimates of the coefficients
$\gammars{2,\,\infty}_{m,\,n}$ to compute $\kappad^{(2,\infty)}$: we
have access to only a few of them and they do not appear to decay fast
enough with $m+n$ so they could be ignored. We must somehow account
for all these missing coefficients if our result is to be reliable.

It is nevertheless possible to design a transparent fitting procedure
which provides results whose accuracy can be easily tested. To this
end, we propose to think of $\kappad^{(2,\,s)}$ as approaching its
asymptotic value by decrements which asymptotically fall on a power
law of the type used above, $b \, s^{-c}$, and extract the asymptotic
value $\kappad^{(2,\,\infty)}$ after estimating the parameters $b$ and
$c$ as functions of $s$.

\begin{figure}[htb]
  \centering
  \begin{subfigure}[t]{0.495\linewidth}
    \includegraphics[width=\textwidth]{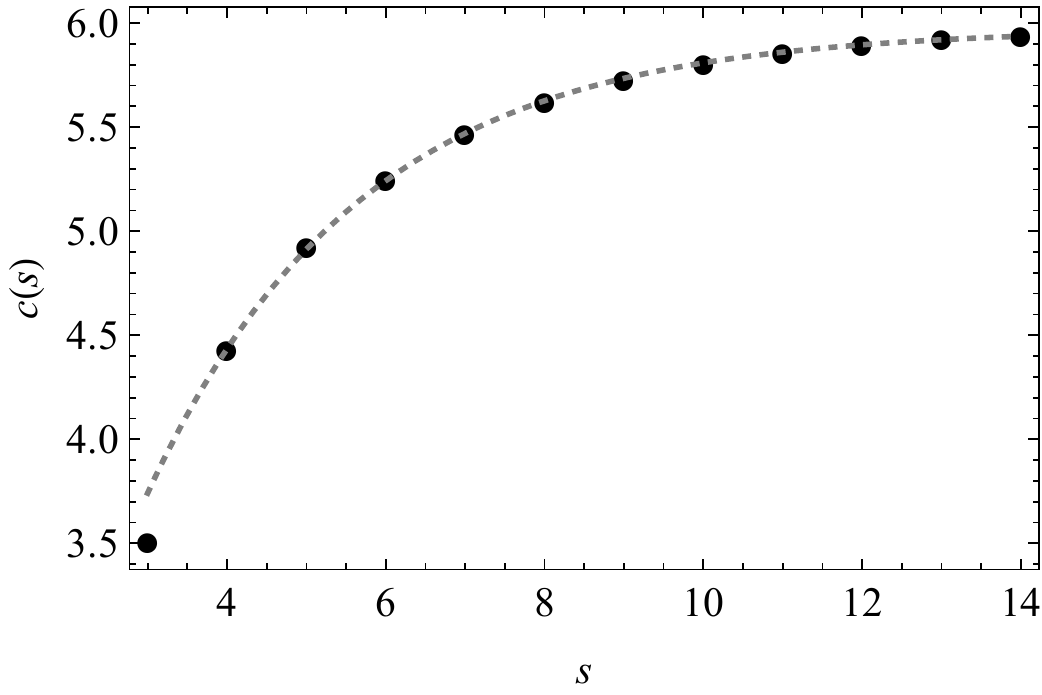}
    \caption{$c(s)$, equation~\eqref{eq:kappadr2_modelincrements_cs}}
    \label{fig:kappadr2_cs}
  \end{subfigure}
  \begin{subfigure}[t]{0.495\linewidth}
    \includegraphics[width=\textwidth]{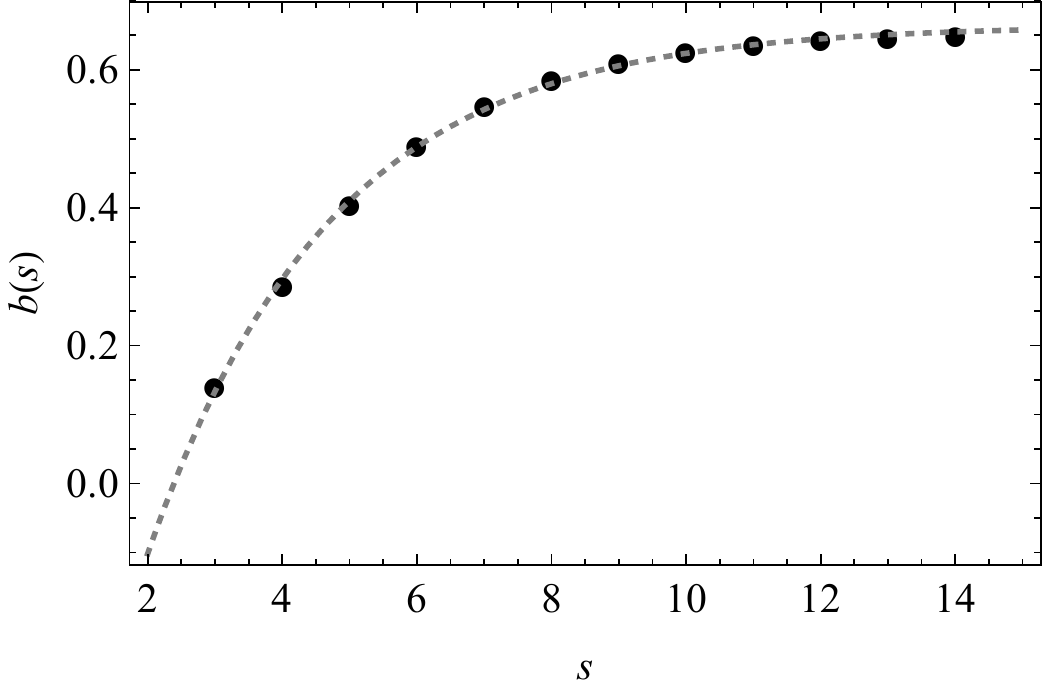}
    \caption{$b(s)$, equation~\eqref{eq:kappadr2_modelincrements_bs}}
    \label{fig:kappadr2_bs}
  \end{subfigure}
  \caption{Exponential convergence of the fitting parameters $c(s)$,
    equation~\eqref{eq:kappadr2_modelincrements_cs}, and $b(s$), equation
    \eqref{eq:kappadr2_modelincrements_bs}. The dotted curves show the
    respective fitting results \eqref{eq:kappadr2_cs} and \eqref{eq:kappadr2_bs}.
  } 
  \label{fig:kappadr2_fit}  
\end{figure}

The asymptotic exponent value is extracted from the computed data
points by considering the decrements $\kappad^{(2,\,s-1)} -
\kappad^{(2,\,s)}$, as plotted in the inset of \fref{fig:kappadr2},
and computing 
\begin{equation}
  \label{eq:kappadr2_modelincrements_cs}
  c(s) = 
  \log \frac{\kappad^{(2,\,s)} - \kappad^{(2,\,s+1)}}
  {\kappad^{(2,\,s-1)} - \kappad^{(2,\,s)}}
  \Big/ \log \frac{s}{s+1}
  \,.
\end{equation}
The result is a sequence of $s$-dependent values $c(s)$ 
plotted in \fref{fig:kappadr2_cs}, which can be seen to converge
exponentially fast to its asymptotic value,
\begin{equation}
  \label{eq:kappadr2_cs}
  c(s) \approx 5.974 \pm 0.005 - (6.8 \pm 0.1) \rme^{-(0.372 \pm 0.004) s}\,.
\end{equation}
Coefficients $b(s)$ are then obtained by solving
\begin{equation}
  \label{eq:kappadr2_modelincrements_bs}
  b(s) = (\kappad^{(2,\,s-1)} - \kappad^{(2,\,s)}) s^{c(\infty)}\,.
\end{equation}
The result is another sequence of $s$-dependent values $b(s)$ plotted
in \fref{fig:kappadr2_bs}, which also displays exponential convergence 
to its asymptotic value,
\begin{equation}
  \label{eq:kappadr2_bs}
  b(s) \approx 0.664 \pm 0.004 - (1.60 \pm 0.05) \rme^{-(0.37 \pm 0.01) s}\,.
\end{equation}
With these quantities, we finally obtain the sought after estimated
value of the dynamic contribution to the heat conductivity due
to bivariate functions:
\begin{equation}
  \label{eq:kappadr2_sinfty}
  \kappad^{(2,\,\infty)} \approx \kappad^{(2,\,15)} - \sum_{s =
    16}^{\infty} b(s) s^{-c(s)}
  = -3.197\,13 \pm 5\times 10^{-5}\,.
\end{equation}
This is the value reported at the bottom of the second column in
table~\ref{tab:gammar2} and the height of the horizontal dotted line
shown in \fref{fig:kappadr2}. The error estimate is inferred from the
data, by  comparing $\kappad^{(2,\,15)}$ and $\kappad^{(2,\,14)} - 
b(15) s^{-c(15)}$. The accuracy of our result is a reflection of the
largest computed degree, $s= 15$, and the smallness of the gap
$\kappad^{(2,\,15)}  
- \kappad^{(2,\,\infty)}$.

Although lower than the best reported upper bound,
$\kappad^{(2,\,15)}$, the estimate \eqref{eq:kappadr2_sinfty} is
restricted to bivariate functions and is therefore not optimal. We can
safely assume $\kappad < \kappad^{(2,\infty)}$, which will be
confirmed below, and have to increase the number of variables in the
trial functions to infer a confidence interval for $\kappad =
\lim_{r,\,s\to\infty} \kappad^{(r,\,s)}$.

\subsection{Extension to multivariate trial functions ($r>2$)
\label{sec:Calculation-(r>2)}}

\begin{table}[htb]
  \begin{center}
    \begin{tabular}{|c|c|c|c|c|c|c|c|c|c|c|c|} 
      \hline 
      $r$ & $s$ & $-10^{4}\kappad^{(r,\,s)}$ & 
      $r$ & $s$ & $-10^{4}\kappad^{(r,\,s)}$ & 
      $r$ & $s$ & $-10^{4}\kappad^{(r,\,s)}$ &
      $r$ & $s$ & $-10^{4}\kappad^{(r,\,s)}$ \\
      \hline 
      \multirow{13}{*}{$3$} & $3$ & $2.247\,93$ &
      \multirow{13}{*}{$4$} & $3$ & $2.299\,71$ &
      \multirow{13}{*}{$5$} & $3$ & $2.313\,40$ &
      \multirow{13}{*}{$6$} & $3$ & $2.317\,84$ 
      \\ 
      \cline{2-3}
      \cline{5-6}
      \cline{8-9}
      \cline{11-12}
      & $4$ & $3.066\,36$ & 
      & $4$ & $3.136\,22$ &
      & $4$ & $3.154\,58$ & 
      & $4$ & $3.160\,52$
      \\ 
      \cline{2-3}
      \cline{5-6}
      \cline{8-9}
      \cline{11-12}
      & $5$ & $3.370\,87$ &
      & $5$ & $3.447\,44$ &
      & $5$ & $3.467\,51$ &    
      & $5$ & $3.473\,99$
      \\
      \cline{2-3}
      \cline{5-6}
      \cline{8-9}
      \cline{11-12}
      & $6$ & $3.494\,82$ &
      & $6$ & $3.574\,13$ &
      & $6$ & $3.594\,90$ &
      & $6$ & $3.601\,61$ 
      \\
      \cline{2-3}
      \cline{5-6}
      \cline{8-9}
      \cline{11-12}
      & $7$ & $3.549\,96$ &
      & $7$ & $3.630\,50$ &
      & $7$ & $3.651\,59$ &
      & $7$ & $3.658\,40$ 
      \\
      \cline{2-3}
      \cline{5-6}
      \cline{8-9}
      \cline{11-12}
      & $8$ & $3.576\,50$ & 
      & $8$ & $3.657\,64$ &
      & $8$ & $3.678\,88$ & 
      & $8$ & $3.685\,73$ 
      \\ 
      \cline{2-3}
      \cline{5-6}
      \cline{8-9}
      \cline{11-12}
      & $9$ & $3.590\,17$ &
      & $9$ & $3.671\,61$ &
      & $9$ & $3.692\,94$ &
      & $9$ & $3.699\,81$ 
      \\ 
      \cline{2-3}
      \cline{5-6}
      \cline{8-9}
      \cline{11-12}
      & $10$ & $3.597\,63$ &
      & $10$ & $3.679\,24$ &
      & $10$ & $3.700\,61$ &
      & $10$ & $3.707\,50$ 
      \\ 
      \cline{2-3}
      \cline{5-6}
      \cline{8-9}
      \cline{11-12}
      & $11$ & $3.601\,92$ &
      & $11$ & $3.683\,62$ &
      & $11$ & $3.705\,02$ &
      & $11$ & $3.711\,91$ 
      \\ 
      \cline{2-3}
      \cline{5-6}
      \cline{8-9}
      \cline{11-12}
      & $12$ & $3.604\,49$ &
      & $12$ & $3.686\,25$ &
      & $12$ & $3.707\,65$ &
      & $12$ & $3.714\,55$ 
      \\ 
      \cline{2-3}
      \cline{5-6}
      \cline{8-9}
      \cline{11-12}
      & $13$ & $3.606\,09$ &
      & $13$ & $3.687\,88$ &
      & $13$ & $3.709\,30$ &
      & $13$ & $3.716\,20$ 
      \\ 
      \cline{2-3}
      \cline{5-6}
      \cline{8-9}
      \cline{11-12}
      & $14$ & $3.607\,12$ &
      & $14$ & $3.688\,93$ &
      & $14$ & $3.710\,35$ &
      &  &  
      \\ 
      \cline{2-3}
      \cline{5-6}
      \cline{8-9}
      & $15$ & $3.607\,80$ &
      & $15$ & $3.689\,63$ &
      &  &  &
      &  &  
      \\ 
      \hline
      $3$ & $\infty$ & $3.609\,61(6)$ & 
      $4$ & $\infty$ & $3.691\,48(6)$ &
      $5$ & $\infty$ & $3.713\,0(1)$ &
      $6$ & $\infty$ & $3.720\,1(3)$  
      \\ 
      \hline 
      \hline 
      \multirow{9}{*}{$7$} & $3$ & $2.319\,52$ &
      \multirow{9}{*}{$8$} & $3$ & $2.320\,24$ &
      \multirow{9}{*}{$9$} & $3$ & $2.320\,58$ &
      \multirow{9}{*}{$10$} & $3$ & $2.320\,75$ 
      \\ 
      \cline{2-3}
      \cline{5-6}
      \cline{8-9}
      \cline{11-12}
      & $4$ & $3.162\,77$ &
      & $4$ & $3.163\,73$ &
      & $4$ & $3.164\,18$ &
      & $4$ & $3.164\,41$ 
      \\ 
      \cline{2-3}
      \cline{5-6}
      \cline{8-9}
      \cline{11-12}
      & $5$ & $3.476\,44$ &
      & $5$ & $3.477\,49$ &
      & $5$ & $3.477\,98$ &
      & $5$ & $3.478\,22$ 
      \\ 
      \cline{2-3}
      \cline{5-6}
      \cline{8-9}
      \cline{11-12}
      & $6$ & $3.604\,14$ &
      & $6$ & $3.605\,22$ &
      & $6$ & $3.605\,72$ &
      & $6$ & $3.605\,98$ 
      \\ 
      \cline{2-3}
      \cline{5-6}
      \cline{8-9}
      \cline{11-12}
      & $7$ & $3.660\,96$ &
      & $7$ & $3.662\,06$ &
      & $7$ & $3.662\,57$ &
      & $7$ & $3.662\,82$ 
      \\ 
      \cline{2-3}
      \cline{5-6}
      \cline{8-9}
      \cline{11-12}
      & $8$ & $3.688\,31$ &
      & $8$ & $3.689\,41$ &
      & $8$ & $3.689\,93$ &
      & $8$ & $3.690\,18$ 
      \\ 
      \cline{2-3}
      \cline{5-6}
      \cline{8-9}
      \cline{11-12}
      & $9$ & $3.702\,40$ &
      & $9$ & $3.703\,50$ &
      & $9$ & $3.704\,02$ &
      &  &  
      \\ 
      \cline{2-3}
      \cline{5-6}
      \cline{8-9}
      & $10$ & $3.710\,09$ &
      & $10$ & $3.711\,20$ &
      &  &  &
      &  &  
      \\  
      \cline{2-3}
      \cline{5-6}
      & $11$ & $3.714\,51$ &
      &  &  &
      &  &  &
      &  &  
      \\  
     \hline
      $7$ & $\infty$ & $3.724(2)$ & 
      $8$ & $\infty$ & $3.727(4)$ &
      $9$ & $\infty$ & $3.73(1)$ &
      $10$ & $\infty$ & $3.74(3)$ 
      \\
      \hline
    \end{tabular}
  \end{center}
  \caption{Dynamic contributions to the heat conductivity obtained
    from the variational formula using the trial functions~\eqref{eq:fr}
    with $r$ variables and maximal degree $\sum_{i=1}^r n_i\leq s$. 
    \label{tab:gammar>2}
  } 
\end{table}

To improve the upper bound \eqref{eq:kappadr2_sinfty} on the dynamical
contribution to the heat conductivity \eqref{eq:kappa_var}, we must go 
beyond bivariate trial functions and transpose the calculations
presented in \sref{sec:Calculation-(r=2)} to multivariate functions of
order $r>2$. For the sake of compressing notations, given the integers
$i\leq j$, we let $\underline{c}_{i:j}$ denote the sequence of indices 
$c_{i}, \dots, c_{j}$ and $\overline{c}$ their sum, $c_{i}+ \dots+
c_{j}$ (we omit the indices). For functions of arbitrary number of 
variables $r$, the variational formula \eqref{eq:kappa_var_r2} thus
becomes  
\begin{multline} 
  \label{eq:kappa_var_r>2}
  \kappad^{(r,\,s)}
  = \mathrm{inf}_{\{\gammars{r,\,s}_{\underline{c}_{1:r}}\}}
    \Bigg[ \sqrt{\frac{3}{2}} \sum_{\substack{m,n = 1\\m+n\leq s}}^{s}
    \gammars{r,\,s}_{m,n} \, A_{m,n,1,0} 
    + 
    \sum_{\substack{\underline{c}_{1:r-1} = 0 \\
        \overline{c} \leq s}}^{s}
    \sum_{m,p = 0}^{s - \overline{c}}
    \gammars{r,\,s}_{\underline{c}_{1:r-1},m} \, 
    \gammars{r,\,s}_{\underline{c}_{1:r-1},p} 
    A_{m,0,p,0} 
    \cr
    + 
    2\sum_{k = 2}^{r} 
    \sum_{\substack{\underline{c}_{1:k-2} = 0 \\
        \overline{c}\leq s}}^{s}
    \sum_{\substack{n,p,q = 0\\
      p+q\leq s - \overline{c}}}^{s - \overline{c}}
    \frac{
      \gammars{r,\,s}_{n,\underline{c}_{1:k-2}} \, 
      \gammars{r,\,s}_{p,q,\underline{c}_{1:k-2}} 
    }
    {(r - k + 1)(r - k + 2)}
    A_{0,n,p,q} 
   +
    \frac{1}{2} \sum_{k = 2}^{r}  \sum_{l = 2}^{k} (2 - \delta_{k,l})
    \cr
    \times
    \sum_{\substack{\underline{c}_{1:r-k+l-2} = 0 \\
        \overline{c}\leq s}}^{s}
    \sum_{\substack{m,n,p,q = 0\\
        m + n \,\&\, p + q \leq s - \overline{c}}}^{s - \overline{c}}
    \frac{
      \gammars{r,\,s}_{\underline{c}_{1:r-k},m,n,\underline{c}_{r-k+1:r-k+l-2}}
      \gammars{r,\,s}_{\underline{c}_{1:r-k},p,q,\underline{c}_{r-k+1:r-k+l-2}}
    }
    {(k-l+1)^2}
    \, 
    A_{m,n,p,q} 
    \Bigg]\,.
\end{multline}
In this expression, we have concatenated index sequences ending or
beginning by sets $\underline{0}_l$ composed of $l$ successive $0$
according to 
\begin{equation}
  \label{eq:gammaconcatenation}
  \gammars{r,\,s}_{\underline{c}_k,\underline{0}_{l}} = 
  \gammars{r,\,s}_{\underline{0}_{l},\underline{c}_k} =
  \frac{r - k - l + 1}{r - k + 1} \gammars{r,\,s}_{\underline{c}_k}\,,
\end{equation}
see \eqref{eq:gammars}, and assumed antisymmetry with respect to
reversing the order of indices,
\begin{equation}
  \label{eq:gammasymmetry}
  \gammars{r,\,s}_{c_{1},\dots,c_{r}} = 
  - \gammars{r,\,s}_{c_{r},\dots,c_{1}}
  \,,
\end{equation}
which also implies that coefficients with a single non-zero index must
vanish, $\gammars{r,\,s}_{c} \equiv 0$. 
As summations over the indices are performed in equation
\eqref{eq:kappa_var_r>2}, further simplifications involving equations
\eqref{eq:gammaconcatenation} and \eqref{eq:gammasymmetry} arise.

Approximations $\kappad^{(r,\,s)}$ obtained by
restricting the computation of the infimum in the variational formula
\eqref{eq:kappa_var_r>2} to multivariate polynomials of order $r$ and
degree $s$ are reported in table~\ref{tab:gammar>2}. 

Each one of the values listed in table~\ref{tab:gammar>2} thus
provides an analytically obtained upper bound on the actual dynamic
contribution to the heat conductivity; more precisely, each such upper
bound is a rational number which we report in decimal approximation to
six significant digits. We may therefore conclude:
\begin{equation} 
  \kappad < - 0.000\,371\,620 \, ,
  \label{eq:var-th} 
\end{equation}
which is obtained for $r=6$ and $s=13$. For this pair of parameters,
the number of coefficients involved in the search of the infimum is
close to $11\,000$. It is about the same number for $r=7$ and
$s=11$. Such large numbers of coefficients set a bound for every $r$
on the degree $s$ for which values $\kappad^{(r,\,s)}$ are within
reach of our computation and thus leaves out empty cells in our table. 

The convergence to an asymptotic value $\kappad^{(r,\,\infty)}$
is observed in every column of table~\ref{tab:gammar>2} as $s$
increases. We can therefore repeat the analysis presented in 
\sref{sec:Calculation-(r=2)} for $r=2$ and extend it to every value of
$r = 3,\dots,\,10$. The accuracy of our scheme to extrapolate to
$s\to\infty$ for a given order $r$ will be tested by the reduced
largest degrees for which values of $\kappad^{(r,\,s)}$ were computed
in the last columns of the table. 

\begin{figure}[htb]
  \centering
  \begin{subfigure}[t]{0.495\linewidth}
    \includegraphics[width=\textwidth]
    {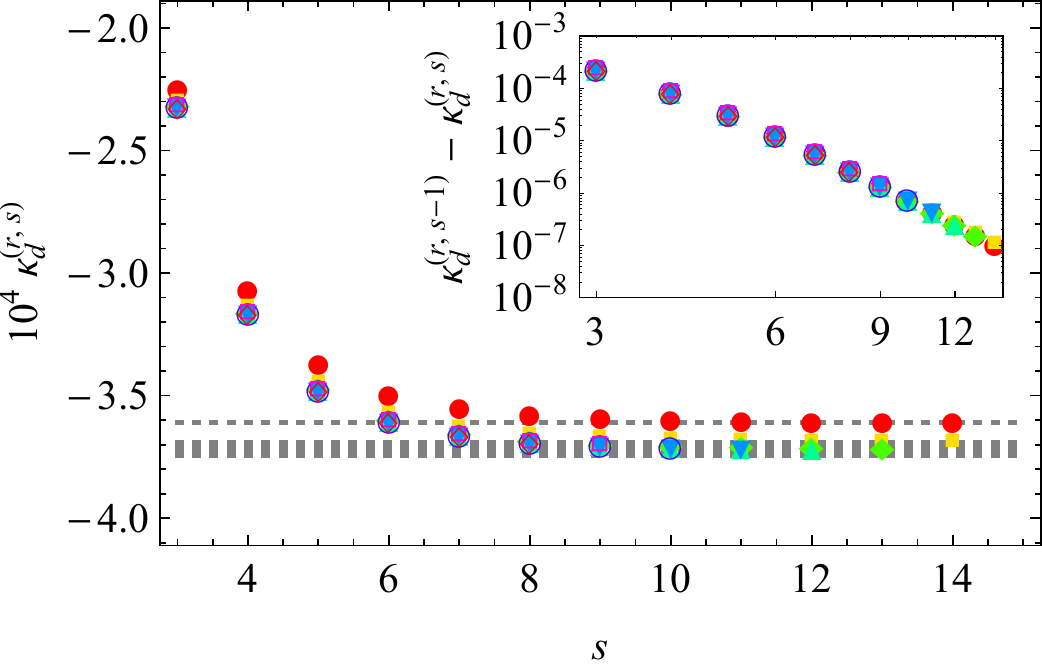}
    \caption{$\kappad^{(r,\,s)}$ v. $s$ \quad ($3 \leq r \leq 10$)}
    \vspace{0.5cm}
    \label{fig:kappadr>2}
  \end{subfigure}
  \begin{subfigure}[t]{0.495\linewidth}
    \includegraphics[width=\textwidth]
    {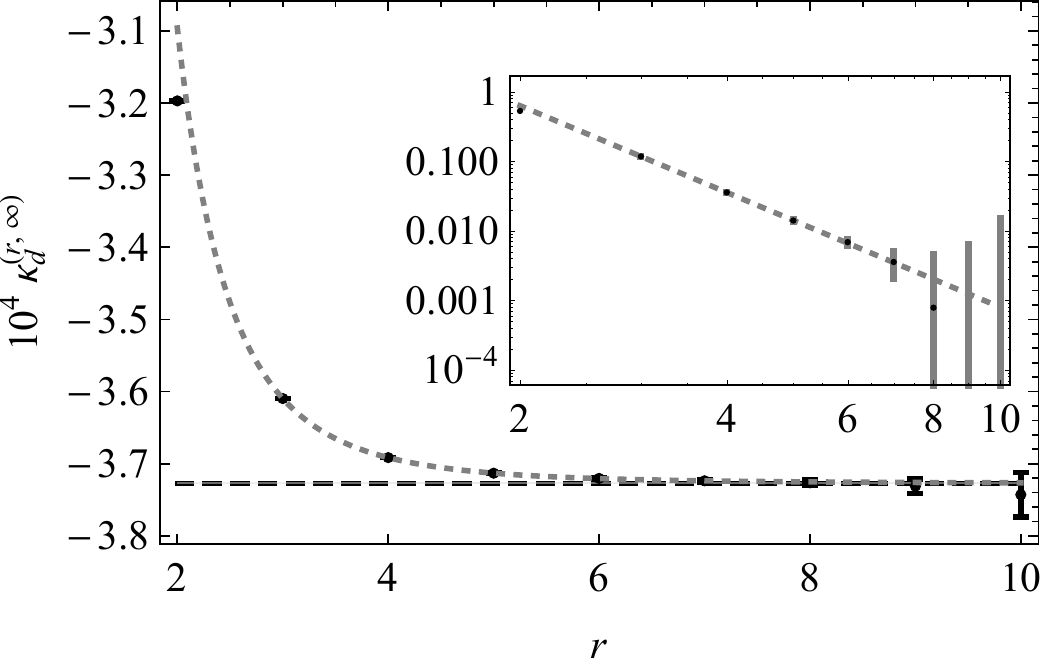}
    \caption{$\kappad^{(r,\infty)}$ v. $r$}
    \label{fig:kappad_sinfty}
  \end{subfigure}
  \caption{(a) Graphical representation of the upper bounds
    $\kappad^{(r,\,s)}$ on the dynamical contribution to the heat
    conductivity reported in table~\ref{tab:gammar>2} for $r = 3,
    \dots,\, 10$. (b) Estimate of $\kappad = 
    \lim_{r\to\infty} \kappad^{(r,\infty)}$ by the nonlinear power-law
    fit \eqref{eq:fitkappad_rinfty}. In both panels, the dotted
    straight lines indicate the results of the extrapolations
    $s\to\infty$ (a) and $r\to\infty$ (b).
  }
  \label{fig:fit_r>2}
\end{figure}

Considering \fref{fig:kappadr>2}, we notice in the inset the
remarkable fact that the decrements $\kappad^{(r,\,s-1)} -
\kappad^{(r,\,s)}$, $3 \leq r \leq 10$, appear to fall along the
same curve as that observed in the inset of \fref{fig:kappadr2}. The
implication is that the exponent $c(s)$, defined in analogy to
equation~\eqref{eq:kappadr2_modelincrements_cs}, must be the same
function of $s$ for every $r$. By accumulating all data points, we find an 
improved fitting curve \eqref{eq:kappadr2_cs},
\begin{equation}
  \label{eq:kappadr>2_cs}
  c(s) \approx 5.989 \pm 0.003 - (6.86 \pm 0.06) \rme^{-(0.371 \pm 0.002) s}\,.
\end{equation}
The coefficients $b_{r}(s)$ are obtained in analogy to equation
\eqref{eq:kappadr2_modelincrements_bs}, 
\begin{align}
  \label{eq:kappadr>2_bs}
  b_{2}(s) 
  & \approx 
    0.692 \pm 0.004 - (1.62 \pm 0.05) \rme^{-(0.36 \pm 0.01) s}\,,
    \cr
    b_{3}(s) 
  & \approx
    0.774 \pm 0.005 - (1.84 \pm 0.07) \rme^{-(0.36 \pm 0.01) s}\,,
    \cr
    b_{4}(s) 
  & \approx
    0.791 \pm 0.006 - (1.88 \pm 0.07) \rme^{-(0.36 \pm 0.01) s}\,,
    \cr
    b_{5}(s) 
  & \approx
    0.800 \pm 0.007 - (1.87 \pm 0.07) \rme^{-(0.36 \pm 0.01) s}\,,
    \cr
    b_{6}(s) 
  & \approx
    0.808 \pm 0.009 - (1.84 \pm 0.07) \rme^{-(0.35 \pm 0.01) s}\,,
    \cr
    b_{7}(s) 
  & \approx
    0.82 \pm 0.01 - (1.81 \pm 0.07) \rme^{-(0.34 \pm 0.01) s}\,,
    \cr
    b_{8}(s) 
  & \approx
    0.83 \pm 0.02 - (1.76 \pm 0.06) \rme^{-(0.33 \pm 0.02) s}\,,
    \cr
    b_{9}(s) 
  & \approx
    0.85 \pm 0.02 - (1.74 \pm 0.06) \rme^{-(0.31 \pm 0.02) s}\,,
    \cr
    b_{10}(s) 
  & \approx
    0.89 \pm 0.03 - (1.70 \pm 0.05) \rme^{-(0.29 \pm 0.02) s}\,.
\end{align}
They display the same kind of exponential convergence to their
asymptotic values as observed in \fref{fig:kappadr2_bs}. For the sake
of comparison, we included here $r=2$, which can be set side-by-side
with equation \eqref{eq:kappadr2_bs}.

With these quantities, we proceed in analogy to equation
\eqref{eq:kappadr2_sinfty} to obtain estimates of the dynamic 
contributions to the heat conductivity due to $r$-variate functions,
$\kappad^{(r,\,\infty)}$. The values are reported in
table~\ref{tab:gammar>2}. They correspond to the heights of the
horizontal dotted lines shown  in \fref{fig:kappadr>2} and are the
data points of \fref{fig:kappad_sinfty} on which the $r\to\infty$
extrapolation is based. The error estimates are inferred from the
data, by  comparing $\kappad^{(r,\,s_{\mathrm{max}})}$ and
$\kappad^{(r,\,s_{\mathrm{max}-1})} -  b_{r}(s_{\mathrm{max}}) 
s^{-c(s_{\mathrm{max}})}$, where, for each $r$, $s_{\mathrm{max}}$ is
the largest degree $s$ for which $\kappad^{(r,\,s)}$ was
computed. Here we note that the differences between the parameter
$b_{2}(s)$ in equation~\eqref{eq:kappadr>2_bs} and $b(s)$ in equation
\eqref{eq:kappadr2_bs} would have the effect of changing the last
digit reported in $\kappad^{(2,\,\infty)}$ by one unit less, which is
well within the corresponding error estimate (which remains
unchanged).

Next we turn to the $r\to\infty$ extrapolation of
$\kappad^{(r,\,\infty)}$, which is the dynamic contribution to the
heat conductivity \eqref{eq:kappad}. The values of the estimated
contributions from $r$-variate functions are reported in
\fref{fig:kappad_sinfty}. As seen from the inset the decrements
$\kappad^{(r-1,\,\infty)} - \kappad^{(r,\,\infty)}$ appear to follow a
power law whose exponent is between $-4$ and $-5$, sufficiently
different from an integer value that we have to resort to a nonlinear
regression of the model. The result of this fit, which excludes $r=2$,
yields 
\begin{equation}
  \label{eq:fitkappad_rinfty}
  \kappad^{(r,\,\infty)}
  \approx -0.000\,372\,72 \pm 6 \times 10^{-8} 
  + (0.001\,12 \pm 5 \times 10^{-5} )\,r^{-4.14 \pm 0.04}\,.
\end{equation}
It is shown as the dotted curve in \fref{fig:kappad_sinfty} (as well as the
inset for the algebraic decay); the dashed horizontal line is the
asymptotic value, $\kappad^{(\infty,\,\infty)}$. The $95\%$ confidence
interval of the first parameter gives our best estimate of the
dynamical contribution to equation~\eqref{eq:kappa_var} ,   
\begin{equation}
  \label{eq:kappadrinftysinfty}
  -0.000\,372\,87 < \kappad < -0.000\,372\,58\,.
\end{equation}
The inferred estimated value of the dynamic contribution $\kappad =
-0.000\,372\,72 (6)$, with seven significant decimals, is consistent with
the upper bound \eqref{eq:var-th}.

\begin{figure}[htb]
  \centering
  \includegraphics[width=0.6\textwidth]
  {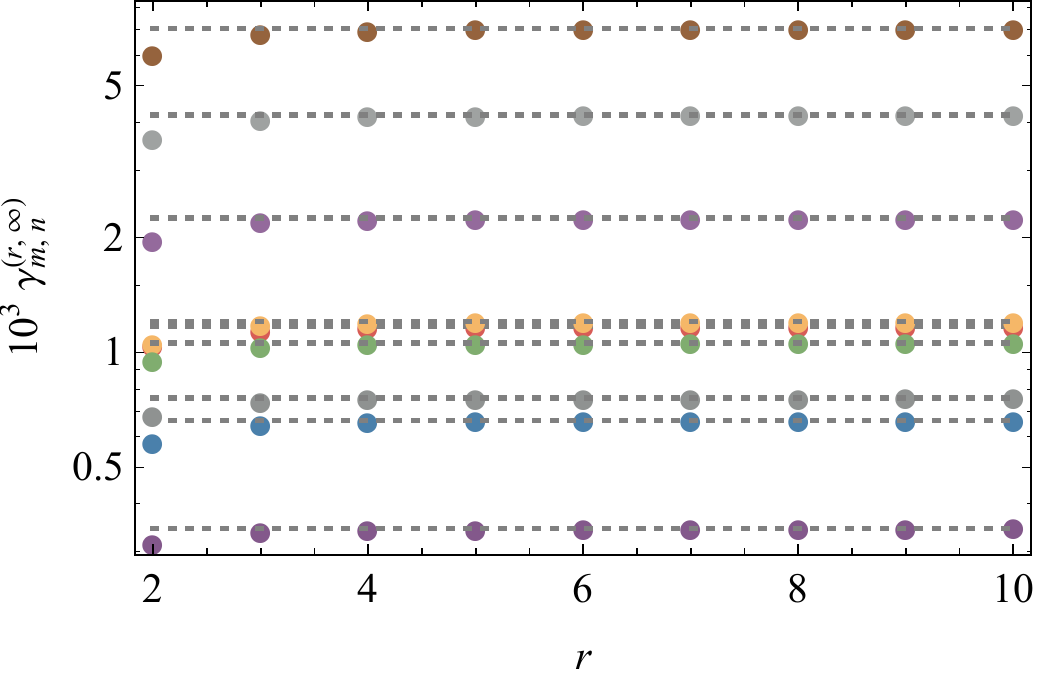}
  \caption{Graphical representation on a semi-log scale of the
    extrapolated coefficients $\gammars{r,\,\infty}_{m,\,n}$ as
    functions of $r$. The straight lines show the $r\to\infty$
    estimates, equation \eqref{eq:gammar>2}.
  }
  \label{fig:gammar>2}
\end{figure}

Our analysis of the coefficients $\gammars{r,\,s}_{m,\,n}$, equation
\eqref{eq:gammar2_model} carries over to $r>2$. The values obtained by the
$s\to\infty$ extrapolation are shown graphically in
\fref{fig:gammar>2} for all pairs $\{m,\,n\}$ such that $m+n \leq
7$. To estimate the $r\to\infty$ extrapolations, we fit these values
by nonlinear regressions with power laws $\gammars{r,\,\infty}_{m,n} 
\approx a + b \, r^{-c}$. The results (excluding $r=2$) are as follows:
\begin{alignat}{2}
  \label{eq:gammar>2}
  \gammars{\infty,\,\infty}_{2,\,1} &= 0.007\,040\,3 \pm 4\times
  10^{-7}\,, &\qquad
  \gammars{\infty,\,\infty}_{3,\,1} &= 0.004\,184\,3 \pm 1\times
  10^{-7}\,, 
  \cr
  \gammars{\infty,\,\infty}_{4,\,1} &= 0.002\,241\,6 \pm 1\times
  10^{-7}\,, &\qquad
  \gammars{\infty,\,\infty}_{3,\,2} &= 0.001\,164\,46 \pm 3\times
  10^{-8}\,, 
  \cr
  \gammars{\infty,\,\infty}_{5,\,1} &= 0.001\,202\,5 \pm 2\times
  10^{-7}\,, &\qquad
  \gammars{\infty,\,\infty}_{4,\,2} &= 0.001\,054\,7 \pm 2\times
  10^{-7}\,, \cr
  \gammars{\infty,\,\infty}_{6,\,1} &= 0.000\,660\,7 \pm 4\times
  10^{-7}\,, &\qquad
  \gammars{\infty,\,\infty}_{5,\,2} &= 0.000\,756\,3 \pm 5\times
  10^{-7}\,, \cr
  \gammars{\infty,\,\infty}_{4,\,3} &= 0.000\,344\,6 \pm 9\times
  10^{-7}\,.&     
\end{alignat}
The error estimates reported here are those returned by the nonlinear
regression. The respective contributions of these coefficients to the
thermal conductivity \eqref{eq:kappa-ststexplicit} are, up to a minus
sign, $0.000\,262\,4$, $0.000\,072\,2$, $0.000\,017\,1$,
$0.000\,005\,6$, $0.000\,004\,1$, $0.000\,004\,2$, $0.000\,001\,0$,
$0.000\,001\,9$, and $0.000\,000\,5$, whose total, $0.000\,369\,0$,
accounts for about $99\%$ of the conductivity
\eqref{eq:kappadrinftysinfty}.  

It would of course be desirable to improve our computation and go
beyond the limited order and degree values reported 
here. This would allow us to refine  our extrapolation  and check the
validity of the model and the precision of the result. While we have
to leave such considerations to future work, we can turn to
simulations of the nonequilibrium steady state to obtain 
an independent estimate on the dynamic contribution to the heat
conductivity.   

\section{Kinetic Monte Carlo simulations}
\label{sec:Simulations}

The nonequilibrium steady state of the stochastic model can be
simulated following along the lines of Gillespie's kinetic Monte Carlo
algorithm \cite{Gillespie:1976p296}. This yields a numerical
determination of the heat conductivity through Fourier's law
\eqref{eq:Fourier}, which can be compared with the theoretical results
described in \sref{sec:Calculation-(r>2)}.  

The method is an improved version\footnote{We are grateful to Imre
  P\'eter   T\'oth for suggesting these improvements.} of that described in 
references~\cite{Gaspard:200811P021, Gaspard:2009P08020}.
We consider a one-dimensional chain of $N$ cells, with both ends in
contact with thermal reservoirs at different temperatures, which we
take to be $T_{-} = \tfrac{1}{2}$ at cell $-(N+1)/2$ and $T_{+} =
\tfrac{3}{2}$ at cell $(N+1)/2$.  Their energies are thus distributed
according to Gamma distributions of shape parameter $\tfrac{3}{2}$ and scale
parameters $T_{\pm}$. Rather than draw a random energy from these
distributions at large enough (constant) rate to simulate the constant
temperature of the reservoirs, it is more precise (as well as it saves
computer time) to consider the integrated form of the 
kernel \eqref{eq:kernel} with  respect to these distributions, which
yields a thermalized kernel  for the interaction of cells $\pm(N+1)/2$
with the thermostats  at temperatures $T_{\pm}$,
\begin{equation}
  \label{eq:bathkernel}
  w_{T}(e|e+h) = 
  \sqrt{\frac{\pi}{8\, e}}
  \times
  \begin{cases}
    0,
    & h< -e,\\
    \rme^{h/T}\mathrm{erf} \left(\sqrt{\frac{e + h}{T}}\right)
    \,,
    & -e \leq h < 0,\\
    \mathrm{erf}  \left(\sqrt{\frac{e}{T}}\right)
    \,,
    & h\geq 0,
  \end{cases}
\end{equation}
where $\mathrm{erf}$ denotes the error function. A moderate price to
pay for this implementation is the numerical determination of the
amount of energy exchanged with the thermostats by the rejection
method \cite[section 7.3.6]{Press:NR}.

At each Monte Carlo step, the time until the next energy exchange
event and the pair involved (including thermostats) is determined from
a collection of clocks associated with each pair of cells. For each
one of them, the frequency $\nu(e_{n}, e_{n+1})$ specifies the
exponential rate of the random distribution from which the time to the
next interaction is generated. For cells in contact with thermal baths,
this rate is 
\begin{equation}
  \label{eq:bathnu}
  \nu_{T}(e) = \sqrt{\frac{T}{8}} \left[
    \rme^{-e/T} + \sqrt{\frac{\pi \, T}{e}} 
    \left(\frac{1}{2} + \frac{e}{T}\right)
    \mathrm{erf}\left(\sqrt{\frac{e}{T}}\right)
    \right]\,,
\end{equation}
where $T = T_{\pm}$. Whenever a clock rings, a uniformly-distributed
random number is generated, which, by inversion of the partially
integrated kernel, yields the amount of energy exchanged between the
two interacting cells. Their clocks are then renewed, along with those
of the relevant neighbouring pairs. At each step, the energies   
\begin{equation}
  \label{eq:energychain}
  \{e_{-(N-1)/2}, \, e_{-(N-3)/2}, \dots, \, e_{(N-3)/2}, \, e_{(N-1)/2}\}  
\end{equation}
in the $N$ cells are so updated while keeping the temperature of the
thermostats constant.   

\begin{figure}[phtb]
  \centering
  \begin{subfigure}[t]{0.55\textwidth}
    \includegraphics[width=\textwidth]
    {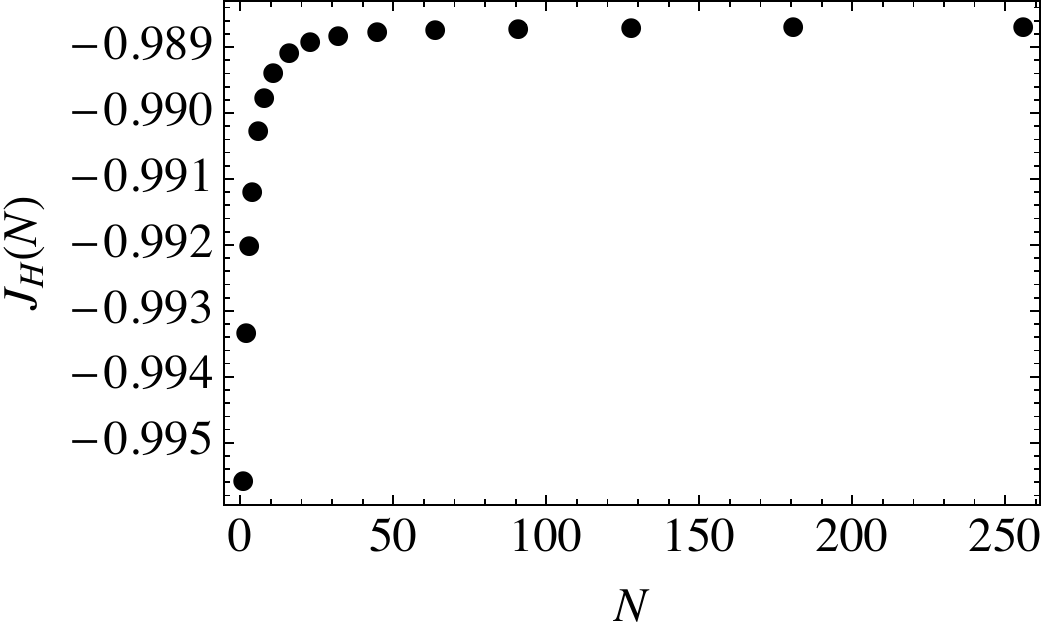}
    \caption{Total current}
    \vspace{0.5cm}
    \label{fig:current}
  \end{subfigure}
  \begin{subfigure}[t]{0.55\textwidth}
    \includegraphics[width=\textwidth]
    {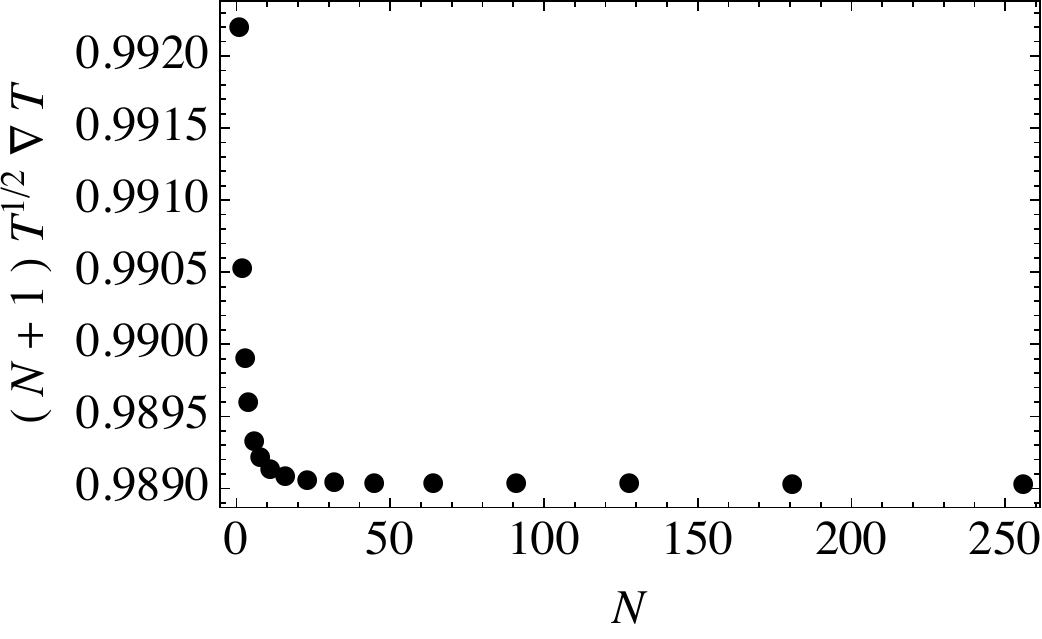}
    \caption{Temperature gradient}
    \vspace{0.5cm}
    \label{fig:temp}
  \end{subfigure}
  \begin{subfigure}[t]{0.55\textwidth}
    \includegraphics[width=\textwidth]
    {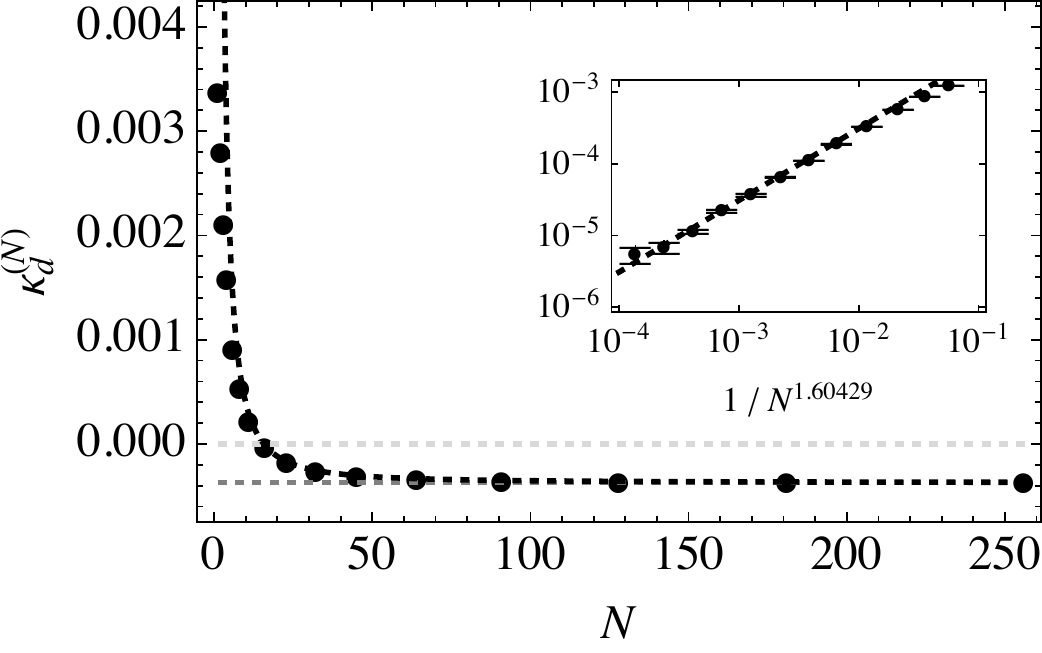}
    \caption{Finite-size dynamic contribution}
    \label{fig:kappaness}
  \end{subfigure}
  \caption{(a) Total current $\Jh$ v. $N$ \eqref{eq:totalcurrent} for
    system sizes of up to $N=256$ cells resulting from the
    nonequilibrium boundary 
    conditions at temperatures $T_{-} = \tfrac{1}{2}$ and
    $T_{+}=\tfrac{3}{2}$. (b) Sum of the local temperature gradient,
    $T_{n+1} - T_{n}$, weighted by the square root of the mean
    temperature $T_{n,\,n+1}$, as it appears in the denominator of
    the right-hand side of equation~\eqref{eq:measurekappaness}. (c)
    Finite-size dynamic contribution to the heat conductivity,
    $\kappad^{(N)}$, and the result of its extrapolation to infinite
    size \eqref{eq:fitkappness}. The inset shows the power-law
    convergence towards the estimated asymptotic value
    $-0.000\,371$.
  }
  \label{fig:currenttemp}
\end{figure}

To measure the average heat flux, we compute the average of the
current \eqref{eq:3dcurrent}, estimating $\avgneq{j(e_{n},
  e_{n+1})}$ between every pair of cells by a time integral. These $N+1$
pairs include the two thermostats, for which we obtain the average
current exchanged with the boundary cell by integrating the current
\eqref{eq:3dcurrent} with respect to the Gamma distribution associated
with the reservoir, 
\begin{equation}
  \label{eq:bathcurrent}
  j_T(e) = \frac{T^{5/2}}{4\sqrt{2}\,e}
  \left[
    \frac{e}{T} \left(\frac{3}{2} - \frac{e}{T}\right) \rme^{-e/T} +
    \sqrt{\pi} \sqrt{\frac{e}{T}}       
    \left(\frac{5}{4} + \frac{e}{T} - \frac{e^2}{T^2}\right)
    \mathrm{erf}\left(\sqrt{\frac{e}{T}}\right)
  \right]\,.
\end{equation}
The average total current, which we denote $\Jh$, is defined as
the sum of all these contributions, 
\begin{equation}
  \label{eq:totalcurrent}
  \Jh(N) = \sum_{n = -(N+1)/2}^{(N-1)/2} \avgneq{j(e_{n}, e_{n+1})}\,.
\end{equation}
Measurements of this quantity are graphically illustrated in
\fref{fig:current} for system sizes ranging from\footnote{The
  exhaustive list is $N = 1, 2, 3, 4, 6, 8, 11, 16, 23, 32, 45, 64,
  91, 128, 181, 256$.} $N=1,\dots,\,256$. 

The heat conductivity is obtained from the above quantity through the
local expression of Fourier's law, $\kappa^{(N)}(T_{n,n+1}) \nabla
T_{n,n+1} = -\avgneq{j(e_{n},   e_{n+1})}$, where $\nabla T_{n,n+1}
\equiv T_{n+1} - T_{n}$ is the difference of local temperatures
between neighbouring cells, which are defined according to
equation~\eqref{eq:localTemp}, and $T_{n,n+1}$ is the arithmetic
average between the two local temperatures. Summing over all cells and
extracting the square-root temperature dependence of the heat 
conductivity, we may thus write  
\begin{equation}
  \label{eq:measurekappaness}
  \frac{\kappa^{(N)}(T)}{\sqrt{T}} = - \frac{\Jh(N)}
  {\sum_{n}\sqrt{T_{n,n+1}} \nabla T_{n,n+1}}
  \,.
\end{equation}
The measured numerator and denominator of the right-hand side of this
equation are separately plotted  in \fref{fig:currenttemp}. By
taking their ratio and subtracting the static contribution, we obtain
the dynamic contribution to the heat conductivity as a function of the
system size; see \fref{fig:kappaness}. An extrapolation to
infinite-system size by a power-law nonlinear fit (excluding
$N\leq45$) yields the result 
\begin{equation}
  \label{eq:fitkappness}
  \kappad^{(N)} \approx  -0.000\,371 \pm 2\times10^{-6} 
  + (0.03 \pm 0.01) N^{-1.6 \pm 0.1} \,.
\end{equation}
The power-law convergence of the data is displayed in the inset of
\fref{fig:kappaness}. The $95\%$ confidence interval of the first
parameter gives the $N\to\infty$ estimate of  the dynamical
contribution to the heat 
conductivity, 
\begin{equation} 
  \label{eq:kappadfromness}
  -0.000\,377 \leq \kappad \leq -0.000\,365 \, .
\end{equation} 
The center of this interval is slighted shifted with respect to the
values found in \sref{sec:Calculation-(r>2)} by application of the
variational formula. Its width is however substantially larger and
contains the confidence interval
\eqref{eq:kappadrinftysinfty}. Moreover the upper bound on the 
right-hand side of equation~\eqref{eq:kappadfromness} is larger  
than the explicit upper bound \eqref{eq:var-th}, which is a reminder
that the interval inferred from Monte Carlo simulations is not as
precise as the found by application of the variational formula. 

We close this section with a brief aside. We argued in \sref{sec:Meso}
that the contribution to the probability density of the steady
state given by equation~\eqref{eq:stst} should be understood as the  
part of the nonequilibrium steady state contributing to the current.
The coefficients
$\gamma_{\dots,0, n_{0}, n_{1}, 0,  \dots} $ of its expansion in terms
of Laguerre polynomials turn out to be antisymmetric with respect to
the exchange of the two indices $n_{0}$ and $n_{1}$, so that the
function which realizes the infimum in the variational formula
\eqref{eq:formula} does not contribute to symmetric observables of two
variables. This however leaves open the possibility that the two-cell
marginal density distribution of the actual steady state may have a
symmetric part the variational formula comes short of revealing. 

\begin{figure}[htb]
  \centering
  \includegraphics[width=0.55\textwidth]
  {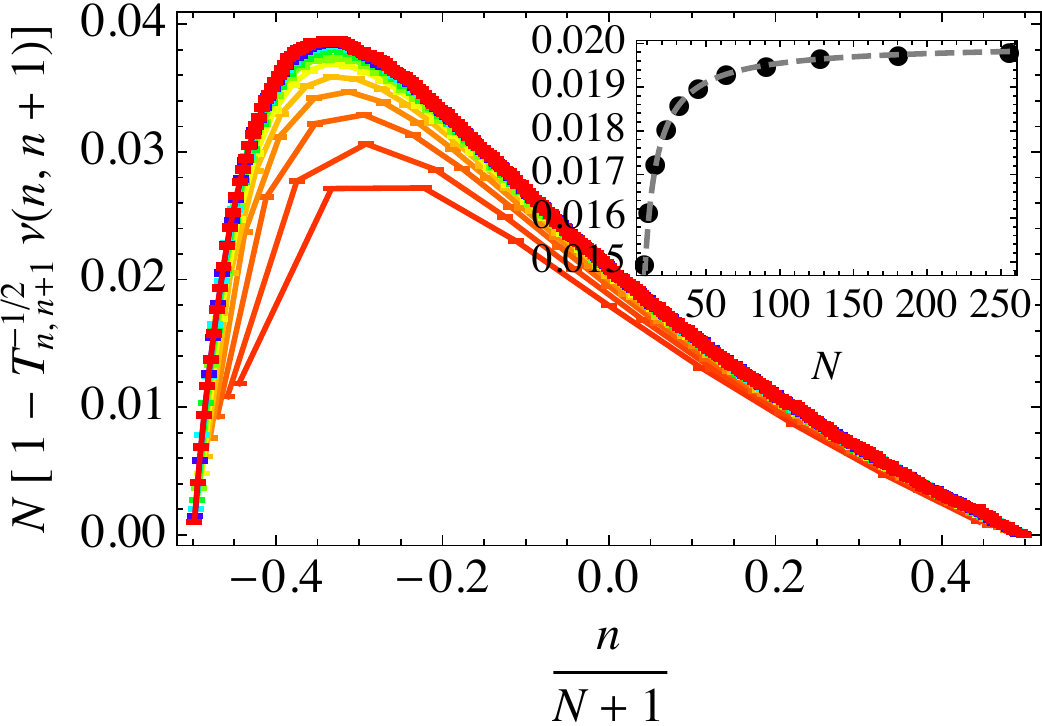}
  \caption{
    Deviations of $\nu(n,\,n+1)$, the measured collision frequency in
    the nonequilibrium steady state, from its local equilibrium
    value, $T_{n,\,n+1}^{1/2}$ multiplied by the system size,
    $N$. Values from $N=8$ to $N=256$ are shown in different
    colors. Inset: average with respect to $n$, $-(N+1)/2 \leq n \leq
    (N-1)/2$, of this quantity as a function of $N$. The dashed line
    shows the power-law fit $0.019\,988 - 0.044\,0766 \, N^{-1}$,
    obtained by nonlinear regression restricted to $N>45$.
  }
  \label{fig:frequencyness}
\end{figure}

The energy exchange frequency \eqref{eq:nu} is such a symmetric
observable. In \fref{fig:frequencyness} we let $\nu(n,\,n+1)$ denote
its ensemble average between cells $n$ and $n+1$ and analyze its
deviations from the local equilibrium contribution $T_{n,\,n+1}^{1/2}$
as the size of the system varies. More precisely, we multiply the
difference $1 - T_{n,\,n+1}^{-1/2} \, \nu(n,\,n+1)$ by $N$ and
estimate the behaviour of this quantity when $N\gg1$. The  
inset of the figure shows how its average converges to an asymptotic
value which our analysis estimates to be in the interval
$[0.019\,922,\,0.020\,054]$ with a power-law convergence in 
$N^{-1}$. We infer from this that symmetric contributions to the
two-cell marginal density distribution of the steady state must
vanish to first order in the gradient expansion\footnote{We now
believe this would have been the correct conclusion of the analysis
presented in reference \cite[section 6]{Gaspard:200811P021}. A 
mistake in the gradient expansion led us to wrongly conclude that
the antisymmetric contributions should vanish.}.

\section{Conclusion}
\label{sec:Conclusion}

We have presented an improved calculation of the heat conductivity of
the energy exchange stochastic model associated with a system of
locally confined hard spheres at the conductor-insulator threshold. 
Sasada's transposition of the variational formula to such a system
\cite{Sasada:2016Thermal, Spohn1990:1227, Spohn:1991book} provides an
efficient instrument to obtain successive exact upper bounds for the
heat transport coefficient, whose values can be extrapolated to infer
an interval of confidence of the dynamic contribution to this 
quantity. Comparisons of this result with values of the heat
conductivity associated with a nonequilibrium steady  
state obtained by kinetic Monte Carlo simulations were presented,
displaying excellent agreement, especially with regards to the
smallness of the numbers reported.

The calculation thus contradicts the conjecture we made earlier in
references~\cite{Gaspard:2008PRL101, Gaspard:2008NJP3004, 
  Gaspard:200811P021, Gaspard:2009P08020, Gaspard:2012p26117} that the
heat conductivity should be equal to the binary collision
frequency. The implication would be that the dynamic contribution to the
conductivity should vanish, as it does in gradient systems
\cite{Spohn:1991book}.  The present results demonstrate that this is
not the case.  The ratio of the heat conductivity and square root of
temperature is indeed slightly smaller than the scaled collision
frequency, with a deviation estimated to be $-0.000\,372\,72(6)$ from
our analysis of the variational formula. Our Monte Carlo simulations
corroborate this result, although with a lesser precision, consistent
to within six significant decimals, $\kappa(T) - \nu(T) \simeq
-0.000\,373\sqrt{T}$. 

The variational formula thus provides a potent tool to compute this
correction. As our results illustrate, the kinetic Monte Carlo
simulations are not as precise. Going to larger system sizes seems to be
necessary, but growing computer times are difficult to manage. This
observation is also reflected by the values of the exponents inferred
from our power-law fits, which are substantially larger (in absolute
value) for the order and degree in the variational formula compared to
that of the system size in the kinetic Monte Carlo simulations. 

Similar results hold for the two-dimensional hard-disc system. In this
case, we obtain the exact upper bound $\kappad < \kappad^{(6,14)} =
-0.000\,893\,56$, which suggests that the dynamical correction to the
heat conductivity is about twice as large for this case than the one
investigated here. Kinetic Monte Carlo simulations using the
stochastic kernel associated with two-dimensional discs are however
much slower as they involve a numerical root-finding algorithm to
determine the amount of energy exchanged when two cells interact. We
suspect that the stochastic energy-exchange process associated with a
system mixing two-dimensional balls and one-dimensional pistons in a
regime of rare interactions \cite{Balint:2016Ballpiston} has a
coefficient of heat conductivity with a dynamic contribution larger
still. However, technical problems due to the mixed nature of this
system have yet to be overcome before the derivation of the
variational formula can be transposed to such systems. 

\ack{The authors are indebted to Makiko Sasada for sharing her
  unpublished results. They wish to acknowledge useful discussions
  with Milton Jara, Carlangelo Liverani, Stefano Olla, Herbert Spohn
  and Domokos Sz\'asz. They also wish to thank Imre P\'eter T\'oth for
  sharing his thoughts on several aspects of this work, and
  specifically with regards to our numerical computations, which his
  insights helped improve substantially. TG wishes to acknowledge the
  hospitality of the Erwin Schr\"odinger Institute, Vienna, on the
  occasion of the conference Hyperbolic Dynamics and Statistical
  Physics held in May 2016, where a preliminary version of this work
  was presented. 
  TG receives financial support from the (Belgian) FRS-FNRS.
  This research was financially supported by the Universit\'e Libre de
  Bruxelles and the Belgian Science Policy Office under the
  Interuniversity Attraction Pole Project P7/18 ``DYGEST''.   
}


\section*{References}


\bibliography{Conductivity.bbl} 

\end{document}